\documentclass[pdftex,11pt,a4paper,parskip=half,headsepline]{scrartcl}

\usepackage[english]{babel}%Sprachen
\usepackage[onehalfspacing]{setspace}%Zeilenabstand
\usepackage[automark]{scrlayer-scrpage}%Kopfzeile
\usepackage[utf8]{inputenc}%Umlaute und Zeichen, für mehr utf8x
\usepackage{latexsym}%weitere Zeichen
\usepackage[T1]{fontenc}%europäischer Schriftsatz
\usepackage{amsmath}
\usepackage{amssymb}
\usepackage{amsfonts}
\usepackage{amstext}
\usepackage{dsfont}
\usepackage{amsthm}%ams Pakete
\usepackage{extarrows}% mehr Pfeile
\usepackage{graphicx}%Grafiken einbinden
\usepackage{xcolor}%mehr Farben
\usepackage{longtable}
\usepackage{enumitem}
\usepackage[longtable]{multirow} %bessere Tabellen
\usepackage{booktabs} % schönere Tabellen
\usepackage{tabularx}
\usepackage{url}%Für Zeilenumbrüche in URLs
\usepackage{todonotes}
\usepackage{algorithm}
\usepackage{algpseudocode}

\usepackage{booktabs} 
\usepackage{siunitx} 
\usepackage{multirow}
\usepackage{hyperref}% für Verweise

\usepackage[backend=biber, bibstyle=apa, maxcitenames=2, citestyle=apa, isbn=false, doi=true, url=false, eprint=true]{biblatex}
\AtEveryBibitem{\clearfield{month}}
\AtEveryCitekey{\clearfield{month}}
\AtEveryBibitem{\clearfield{day}}
\AtEveryCitekey{\clearfield{day}}
\AtEveryBibitem{\clearfield{note}}
\AtEveryCitekey{\clearfield{note}}
\NewBibliographyString{toappear}
\DefineBibliographyStrings{english}{%
  toappear = {to appear in:},
  version = {\textsf{R} package version},
} 
\usepackage{csquotes}

\addto\extrasenglish{

}

\title{Inference for Functional Matched Pairs with Missingness in a Single Arm}
\subtitle{}

\author{Marléne Baumeister$^{1}$, Markus Pauly$^{1,2}$, \L ukasz Smaga$^{\dagger,3}$}

\makeatletter

\clearpairofpagestyles
\ihead{\headmark}
\automark{section}
\ohead{Functional Missingness in One Arm}
\newcolumntype{R}{>{\raggedleft\arraybackslash}X}

\newtheorem{satz}{Theorem}[]

\newtheorem{prop}[satz]{Proposition}
\newtheorem*{ass}{Assumptions}

\newcommand{\br}{\mathbb{R}}
\newcommand{\bn}{\mathbb{N}}

\DeclareMathOperator{\prob}{Pr}
\DeclareMathOperator{\E}{E}
\DeclareMathOperator{\D}{D} %Packing Number

\DeclareMathOperator{\cov}{Cov}

\DeclareMathOperator{\tr}{tr}
\DeclareMathOperator{\GP}{GP}
\DeclareMathOperator{\SP}{SP}

\newcommand{\vertiii}[1]{{\left\vert\kern-0.25ex\left\vert\kern-0.25ex\left\vert #1 \right\vert\kern-0.25ex\right\vert\kern-0.25ex\right\vert}} % spectral norm

\definecolor{bblue}{HTML}{28539E}
\definecolor{turquoise}{HTML}{47C2D6}
\definecolor{pink}{HTML}{E39AB7}
\definecolor{lgreen}{HTML}{94e017 }
\definecolor{orange}{HTML}{e88909}
\definecolor{sagegreen}{RGB}{143, 188, 143}     % Stonowana, modna szałwia

\begin{document}

% Titelseite
\thispagestyle{empty}
\maketitle

% Please change affiliations, it necessary
\renewcommand*{\thefootnote}{\fnsymbol{footnote}}
\footnotetext[2]{Corresponding author: \L ukasz Smaga, \textsf{e-mail: ls@amu.edu.pl}}
\renewcommand*{\thefootnote}{\arabic{footnote}}
\footnotetext[1]{Department of Statistics, TU Dortmund University, Germany}
\footnotetext[2]{Research Center Trustworthy Data Science and Security, UA Ruhr, Germany}
\footnotetext[3]{Faculty of Mathematics and Computer Science, Adam Mickiewicz University in Pozna{\'n}, Poland}

\section*{Abstract} %150-250 words, no abbreviations.
We propose a new statistical test for the equality of mean functions in matched-pairs functional data when entire functions are missing in one arm. 
Unlike complete case analyses, we use the full information from both, the complete and the incomplete pairs. 
Our approach thereby accounts for the dependence between the paired functional observations without imposing a specific covariance structure.
As statistics, we introduce three pointwise quadratic forms and construct global test statistics using supremum or integral functionals. 
Critical values are obtained by a parametric bootstrap procedure. 
Using empirical process theory, we establish that the proposed tests are all of asymptotic level $\alpha$ and consistent against fixed alternatives. 
An extensive simulation study investigates their finite-sample properties, and an application to real data illustrates the practical relevance of the method.

\textbf{Keywords:} %Don't use parts of the title as keywords ;-)
Dependent Functional Samples, Missing Data, Parametric Bootstrap, Qua\-dratic Form-type Test Statistic, Two-Sample Testing.

\section[Introduction]{Introduction and Motivation}\label{sec:intro}
Functional data occur in many scientific fields, including biostatistics or econometrics, as modern measurement technology allow the observation of dense signals or trajectories over time or space \parencite[e.g.,][]{sorensen_introduction_2013,ullah_applications_2013,gertheiss_functional_2024,orozco_functional_2025,cao_statistical_2026}. 
Functional data analysis (\textsc{fda}) provides a natural framework for the analysis of such data, accounting for the inherent dependence structure
\parencite[e.g.,][]{ramsay_functional_2005, horvath_inference_2012, zhang_analysis_2014, crainiceanu_functional_2024}.
An important inference problem in \textsc{fda} is the comparison of mean functions. 
A variety of hypothesis tests has been developed for this purpose, including methods for complex factorial designs \parencite[e.g.,][]{zhang_two-sample_2010, zhang_new_2019, zhang_one-way_2014, jimenez-gamero_testing_2021}, multivariate functional data \parencite[e.g.,][]{gorecki_multivariate_2017, qiu_global_2026, zhu_one-way_2022}, and multiple hypotheses \parencite[e.g.,][]{munko_general_2025, munko_multiple_2025}. 
In this paper, we focus on functional matched-pairs designs, which occurs when two functional measurements of the same subject or experimental unit are collected. 
Here, the dependence between the paired observations must be taken into account when testing for equality of the two mean functions. 
Different approaches are available for this setting. For example, \textcite[][]{smaga_note_2020} proposed integral and supremum-type test statistics based on pointwise standardized differences of the functional sample means and investigated several approaches to compute critical values with a permutation procedure showing the best performance.
More recently, \textcite{Kong_et_al_2026} introduced an $F$-type statistic that uses a global measure of within-condition variability and an $F$-approximation with data-driven degrees of freedom.

Despite these methodological advances, handling missing data in \textsc{fda} remains challenging. 
While \textcite{jang_bayesian_2021} proposed a computationally demanding Bayesian multiple imputation approach for bivariate functional data with missings, our resampling-based inference procedures, \textcite{baumeister_resampling-based_2026} require missing observations in both arms. 
Therefore, there is a lack of inference procedures for functional matched-pairs designs with missngness in one arm only. 
Such one-sided missingness arises naturally in repeated measures designs, e.g. in clincal trials, where subjects are observed at baseline but may drop out of the study before follow-up (see, e.g., Section~\ref{sec:real_data_example}). 
To address this problem, we consider a specific contrast matrix together with a covariance matrix estimator that reflect the dependence structure between the complete and missing functional data parts. 
Based on these estimators, we construct three pointwise quadratic-form test statistics and globalize them using supremum or integral functionals. 
Critical values are computed by a parametric bootstrap procedure. 
By incorporating all available observations, our procedures avoid the loss of information occurring in complete cases analyses and the need to impute missing functional observations. 
This is especially relevant in small sample settings, such as pre- or early clinical studies, where discarding incomplete pairs can drastically reduce the effective sample size.

The paper is structured as follows.
We introduce the statistical framework in Section~\ref{sec:setup}, while Section~\ref{sec:methods} presents the testing procedures and their asymptotic properties. 
The parametric bootstrap procedure is explained in Subsection~\ref{sec:resampling}. 
A detailed simulation study in Section~\ref{sec:simulation} and an illustrative data analysis in Section~\ref{sec:real_data_example} complete the numerical investigations of the paper. 
Finally, Section~\ref{sec:end} concludes with a discussion and an outlook.

\section[Model]{A Model of Matched Pairs for Functional Data}\label{sec:setup}
Let $(\Omega, \mathcal{A}, P)$ be a probability space and $(\mathcal{T}, \zeta)$ be a compact pseudometric space, e.g. $\mathcal{T}=[0,1]$ with the Euclidean metric. 
We consider a two-dimensional stochastic process $\{\mathbf{X}(t)=(X_1(t),X_2(t)):t\in\mathcal{T}\}$,  $\mathbf{X}:\,\Omega\to C^2(\mathcal{T}),\;\boldsymbol{\omega}\mapsto\mathbf{X}(\boldsymbol{\omega}),$ 
taking values in the space $C^2(\mathcal{T})=\{f:\,\mathcal{T}\to\br^2\,|\,f\;\text{continuous}\}$. 
We refer to Appendix~\ref{sec:eptheory} for more technical details. 
Further, we assume that $\mathbf{X}$ has  a mean function $\boldsymbol{\eta}:\mathcal{T}\to\br^2$ and a covariance function $\boldsymbol{\Gamma}:\mathcal{T}^2\to\br^{2\times 2}$.
In the following, the notation $\mathbf{X}\sim \SP_2(\boldsymbol{\eta},\boldsymbol{\Sigma})$ is used to describe the stochastic process $\mathbf{X}$, and analogously $\mathbf{Z}\sim \GP_d(\E(\mathbf{Z}),\cov(\mathbf{Z}))$ denotes a $d$-dimensional Gaussian process whose mean and covariance functions are given by $\E(\mathbf{Z})$ and $\cov(\mathbf{Z})$, respectively.
We consider $n\in\bn$ independent realizations of the stochastic process $\mathbf{X}$:
\begin{align}\label{eq:data}
	\mathbf{x}_1,\dots,\mathbf{x}_n: \mathcal{T}\to\br^2,\, \mathbf{x}_j(t) = (x_{1j}(t), x_{2j}(t))^\top,\,j\in\{1,\dots,n\}.
\end{align}

We assume that entire functions may be missing in one arm only with missingness occurring completely at random (\textsc{mcar}); w.l.o.g., we assume that this is the second arm.
Therefore, the observations in Equation~(\ref{eq:data}) are divided into a complete part of size $n_1$ and an incomplete part (where only the first component is observed) of size $n_2=n-n_1$ for every $t\in\mathcal{T}$. For ease of presentation, we treat $n_1$ and $n_2$ as deterministic, which can be justified by conditioning on the observed sample sizes under \textsc{mcar}. 
The sample can then be represented as
\begin{align}\label{eq:missingdata}
	\mathbf{x}
	=\underbrace{
		\begin{pmatrix}
			x_{11}^{(c)}(t)\\
			x_{21}^{(c)}(t)
		\end{pmatrix},
		\dots,
		\begin{pmatrix}
			x_{1n_1}^{(c)}(t)\\
			x_{2n_1}^{(c)}(t)
		\end{pmatrix} 
	}_{ \mathbf{x}^{(c)}}
	,
	\underbrace{
		\begin{pmatrix}
			x_{11}^{(m)}(t)\\
			---
		\end{pmatrix},
		\dots,
		\begin{pmatrix}
			x_{1n_2}^{(m)}(t)\\
			---
		\end{pmatrix}
	}_ {\mathbf{x}^{(m)}}.
\end{align}
The independence of the observations as realizations of the stochastic process $\mathbf{X}$ implies that the complete part $\mathbf{x}^{(c)}(t)=(\mathbf{x}_{1}^{(c)}(t), \mathbf{x}_{2}^{(c)}(t))^\top$ is independent and identically distributed (i.i.d.) with mean function $\boldsymbol{\eta}(t)=(\eta_1(t),\eta_2(t))^\top$ and covariance matrix
\begin{align}\label{eq:covariance}
	\boldsymbol{\Gamma}(s,t)=
	\begin{pmatrix}
		\gamma_1(s,t) & \gamma_{12}(s,t)\\
		\gamma_{12}(s,t) & \gamma_2(s,t)
	\end{pmatrix},
\end{align}
for every $s,t\in\mathcal{T}$.
In the missing part $\mathbf{x}^{(m)}$, the functions in the first arm $x_{1j}^{(m)}$, $j\in\{1,\dots,n_2\}$, are i.i.d. as well with mean function $\eta_1(t)$ and covariance $\gamma_1(s,t)$, $s,t\in\mathcal{T}$.
This set-up extends the classical non-functional matched-pairs design of \textcite{amro_asymptotic-based_2021} with  missing data in one arm to the functional setting. 

Our aim is to test the hypothesis of equal mean functions
\begin{align}\label{eq:hypothesis}
	\mathcal{H}_0:\,\forall\, t\in\mathcal{T}:\,\eta_1(t)=\eta_2(t),\quad
	\mathcal{H}_1:\,\exists\, t\in\mathcal{T}:\,\eta_1(t)\not=\eta_2(t).
\end{align}
To infer this hypothesis, we define the so-called subject-effect functions $\mathbf{v}_{j}(t)=\mathbf{x}_{j}(t)-\boldsymbol{\eta}(t)$, $j\in\{1,\dots,n\}$, $t\in\mathcal{T}$.
Consequently, they are i.i.d. realizations of a stochastic process $\{\mathbf{V}(t): t\in\mathcal{T}\}$, $\mathbf{V}:\,\Omega\to C^2(\mathcal{T})$ with zero mean function and covariance function $\boldsymbol{\Gamma}$ for every $t\in\mathcal{T}$.
To this end, we impose the following assumptions:
\begin{ass}%\label{thm:assumptions}
	Consider the data defined in Equations~(\ref{eq:data}) and (\ref{eq:missingdata}). 
	\begin{enumerate}[label=(A\arabic*)]
		\item\label{ass:A2} There exist $t^*\in\mathcal{T}$ and $J\ge 2$ such that $\E[\|\mathbf{v}_1(t^*)\|_2^J]<\infty$, where here and throughout $\|\cdot\|_2$ denotes the Euclidean norm.% of a vector of numbers.
		\item\label{ass:A3} The covariance matrix $\boldsymbol{\Gamma}(t,t)$ is positive definite for all $t\in\mathcal{T}$.
		\item\label{ass:A4} The complete and incomplete parts 
		are asymptotically non-negligible, i.e., $n_i/n\rightarrow\kappa_i>0$ for $\min\{n_1,n_2\}\to \infty$, $i\in\{1,2\}$. 
		\item\label{ass:A5} There exists a constant $C\ge0$ such that, for all $s,t\in\mathcal{T}$,  $\E[\|\mathbf{v}_1(t)-\mathbf{v}_1(s)\|_2^J]\le C\cdot\zeta(s,t)^J$, and there exists $\nu>0$ such that
		\begin{align}\label{eq:packingnumber}
			\int_0^{\nu}\D(\epsilon, \zeta)^{2/J}d\epsilon<\infty.
		\end{align}
		Here, $J$ is as in Assumption~\ref{ass:A2} and $\D(\epsilon, \zeta)$ denotes the packing number, see \textcite[][Def. 2.2.3]{van_der_vaart_weak_2023}.
	\end{enumerate}
\end{ass}
For further details of the model, we refer to  \textcite{baumeister_resampling-based_2026}. 
We highlight that the Assumptions~\ref{ass:A2} and~(\ref{ass:A5}) provide the necessary moment and regularity conditions to apply concepts from empirical process theory \parencite{van_der_vaart_weak_2023}. 
Assumption~\ref{ass:A4} ensures that neither the complete nor the incomplete group is asymptotically negligible. 
It is a common assumption in asymptotic inference for general factorial designs, compare \textcite{amro_asymptotic-based_2021} for this specific version. 
Unless stated otherwise, convergence is always meant as $\min\{n_1,n_2\}\to\infty$. 
We write $\stackrel{d}{\longrightarrow}$ and $\stackrel{p}{\longrightarrow}$ to denote convergence in distribution and probability in the sense of \textcite{van_der_vaart_weak_2023}, respectively.
Moreover, $\mathbf{0}_d$ denotes a vector consisting of $d$ zeros.

\section{Statistical Methodology}\label{sec:methods}
A natural way to infer our hypothesis is to consider estimators for the mean difference $\eta_1(t)-\eta_2(t)$ for every $t\in\mathcal{T}$.
In a scenario without missingness, we just calculate empirical means on the whole paired data.
On order to combine information from complete and incomplete pairs in context of missingness, we consider two estimators for $\eta_1(t)-\eta_2(t)$, $t\in\mathcal{T}$. 
The first is based on the complete pairs only, while the second combines observations from incomplete and complete pairs. 
Since both estimators depend on the complete pairs, their dependence needs to be taken into account in the test statistics. 
To this end, we use a proper contrast matrix in combination with covariance matrix estimators as described below.

Based on the complete pairs $\mathbf{x}^{(c)}$, an unbiased estimator for the pooled mean function $\boldsymbol{\eta}: \mathcal{T}\to\br^2$ is given by $\boldsymbol{\hat{\eta}}^{(c)}(t)=(\hat\eta_1^{(c)}(t),\hat\eta_2^{(c)}(t)):=1/n_1\sum_{j=1}^{n_1}\mathbf{x}^{(c)}_{j}(t), \;t\in\mathcal{T}$. 
Similarly, $\eta_1:\mathcal{T}\to\br$ can also be estimated from the incomplete data part $\mathbf{x}^{(m)}$ alone by $\hat{\eta}_1^{(m)}(t):=1/n_2\sum_{j=1}^{n_2}x^{(m)}_{1j}(t)$, $t\in\mathcal{T}$. 
To estimate the mean difference $\eta_1-\eta_2$, we can therefore use $\hat\eta_1^{(c)}(t)- \hat\eta_2^{(c)}(t)$ as well as $\hat\eta_1^{(m)}(t)- \hat\eta_2^{(c)}(t)$ for every $t\in\mathcal{T}$. 
To jointly handle both, we define for every $t\in\mathcal{T}$ the vector 
\begin{align*}
	\boldsymbol{\hat\mu}(t):=\left(\hat{\boldsymbol{\eta}}^{(c)}(t)^\top, \hat\eta_1^{(m)}(t)\right)^\top =\left(\hat\eta_1^{(c)}(t),\hat\eta_2^{(c)}(t), \hat\eta_1^{(m)}(t)\right)^\top
\end{align*}
and the contrast matrix
\begin{align*}
	\mathbf{A}:=\begin{pmatrix}
		1 & -1 & 0 \\
		0 & -1 & 1
	\end{pmatrix}\in\br^{2\times 3}.
\end{align*}
Then, $\mathbf{A}\hat{\boldsymbol{\mu}} = (\hat\eta_1^{(c)}- \hat\eta_2^{(c)},\hat\eta_1^{(m)}- \hat\eta_2^{(c)})$ and the null hypothesis can be written as $\mathcal{H}_0:\mathbf{A}\boldsymbol{\mu}(t)=\mathbf{0}$ for all $t\in\mathcal{T}$, where $\boldsymbol{\mu}(t)=(\boldsymbol{\eta}(t)^\top, \eta_1(t))^\top$.
To account for the covariance structure of $\mathbf{A}\hat{\boldsymbol{\mu}}$, we consider the empirical covariance functions of the complete and incomplete part. 
They are given by 
\begin{align}
	\boldsymbol{\hat\Gamma}^{(c)}(s,t)&=\begin{pmatrix}
		\hat\gamma^{(c)}_1(s,t) & \hat\gamma^{(c)}_{12}(s,t)\nonumber\\
		\hat\gamma_{12}^{(c)}(s,t) & \hat\gamma_2^{(c)}(s,t)
	\end{pmatrix}\\
	&:=\frac{1}{n_1-1}\sum_{j=1}^{n_1}\left(\mathbf{x}^{(c)}_j(s)-\boldsymbol{\hat{\eta}}^{(c)}(s)\right)\left(\mathbf{x}^{(c)}_j(t)-\boldsymbol{\hat{\eta}}^{(c)}(t)\right)^\top,\; s,t\in\mathcal{T},\label{eq:cempcovariance}
\end{align}
and 
\begin{align}\label{eq:mempcovariance}
	\hat\gamma_1^{(m)}(s,t)=\frac{1}{n_2-1}\sum_{j=1}^{n_2}\left(x^{(m)}_{1j}(s)-\hat{\eta}_1^{(m)}(s)\right)\left(x^{(m)}_{1j}(t)-\hat{\eta}_1^{(m)}(t)\right),\; s,t\in\mathcal{T},
\end{align}
respectively. 
Under the Assumptions~\ref{ass:A2} and \ref{ass:A5}, it follows from Lemma 6 in \textcite{munko_multiple_2025}, that $\boldsymbol{\hat\Gamma}^{(c)}(t,t)$  and $\hat\gamma_1^{(m)}(t,t)$ are both uniformly consistent, see also \textcite{baumeister_resampling-based_2026}.
Therefore, the asymptotic distribution of $\sqrt{n}\mathbf{A}(\hat{\boldsymbol{\mu}}-{\boldsymbol{\mu}})$ follows from Theorem~\ref{thm:clt} in Section~\ref{sec:proofs}: 
\begin{prop}\label{thm:Aclt}
	Assume Assumptions~\ref{ass:A4} and~\ref{ass:A5}.
	Then, we have 
	\begin{align*}
		\left(\sqrt{n}\mathbf{A}\left(\boldsymbol{\hat\mu}-\boldsymbol{\mu}\right)(t)\right)_{t\in\mathcal{T}}\stackrel{d}{\longrightarrow}\mathbf{Z}
	\end{align*}
	in $C^2(\mathcal{T})$, where $	\mathbf{Z}\sim \GP_2(\mathbf{0},\cov(\mathbf{Z}))$ with $\cov(\mathbf{Z}(t),\mathbf{Z}(s))
	= \mathbf{A}\boldsymbol{\Sigma}(s,t)\mathbf{A}^\top$ and
	\begin{align*}
		\boldsymbol{\Sigma}(s,t)=\begin{pmatrix}
			\kappa_1^{-1}\boldsymbol{\Gamma}(s,t) & \mathbf{0}_2\\
			\mathbf{0}_2^\top & \kappa_2^{-1}\gamma_1(s,t)
		\end{pmatrix}.
	\end{align*}
\end{prop}
The proof of Proposition~\ref{thm:Aclt} and all other proofs can be found in Section~\ref{sec:proofs}.
The asymptotic covariance $\boldsymbol{\Sigma}(t,t)$ can be estimated by the plug-in estimator 
\begin{align}\label{eq:estim_sigma}
	\boldsymbol{\hat\Sigma}(t,t)= \begin{pmatrix}
		\frac{n}{n_1}\boldsymbol{\hat\Gamma}^{(c)}(t,t) & \mathbf{0}_2\\
		\mathbf{0}_2^\top & \frac{n}{n_2}\hat\gamma_1^{(m)}(t,t)
	\end{pmatrix}.
\end{align}
Its consistency follows from the consistency of the two empirical covariance estimators $\boldsymbol{\hat\Gamma}^{(c)}(t,t)$  and $\hat\gamma_1^{(m)}(t,t)$ and Assumption~\ref{ass:A4}.

Based on this result, we can construct pointwise quadratic form-type test statistics.
Following  \textcite{amro_asymptotic-based_2021}, we propose three versions, which can all be expressed in the form
\begin{align}\label{eq:pw_teststatistic}
	T_n(\mathbf{A},\mathbf{x},t)=n\left(\mathbf{A}\boldsymbol{\hat\mu}(t)\right)^\top\Xi\left(\mathbf{A},\boldsymbol{\hat\Sigma},t\right)\mathbf{A}\boldsymbol{\hat\mu}(t),\,t\in\mathcal{T}.
\end{align}
Here, $\Xi$ is a function of $\boldsymbol{A}$, $\boldsymbol{\hat\Sigma}$ and $t$. 
Its choice determines the type of test statistic: 
For a Wald-type statistic (\textsc{wts}), we use $\Xi_{\textsc{wts}}(\mathbf{A},\boldsymbol{\hat\Sigma},t)= (\mathbf{A}\boldsymbol{\hat\Sigma}(t,t)\mathbf{A}^\top)^+$, where $\mathbf{B}^+$ denotes the Moore-Penrose inverse of a matrix $\mathbf{B}$.
Originally introduced by \textcite{wald_tests_1943}, this type of statistic has been considered in testing procedures with heteroscedastic metric data \parencite[e.g.,][]{konietschke_parametric_2015, pauly_asymptotic_2015}.
An \textsc{anova}-type statistic (\textsc{ats}) is realized by the function $\Xi_{\textsc{ats}}(\mathbf{A},\boldsymbol{\hat\Sigma},t)=\tr(\mathbf{A}\boldsymbol{\hat\Sigma}(t,t)\mathbf{A}^\top)^{-1}$.
This type of statistic was introduced by \textcite{brunner_box-type_1997} and has been applied in factorial \textsc{anova} settings \parencite[e.g.,][]{friedrich_wild_2017, dobler_nonparametric_2020}.
Finally, we define a modified \textsc{anova}-type statistic (\textsc{mats}) by $\Xi_{\textsc{mats}}(\mathbf{A},\boldsymbol{\hat\Sigma},t)=(\mathbf{A}\boldsymbol{\hat\Sigma}(t,t)\mathbf{A}^\top)_0^+$, where $\mathbf{B}_0$ denotes the diagonal matrix having the diagonal elements of $\mathbf{B}$ and zeros elsewhere. 
This type of statistic was introduced by \textcite{friedrich_mats_2018} for \textsc{manova} settings, and was also implemented for other estimands and testing procedures \parencite[e.g.,][]{baumeister_quantile-based_2024, sattler_testing_2022, sattler_quadratic_2026}.

To obtain test statistics for the functional hypothesis in Equation~\eqref{eq:hypothesis}, we globalize the pointwise statistics over $\mathcal{T}$ using either a supremum or an integral functional as is common in \textsc{fda} \parencite[e.g.,][]{baumeister_resampling-based_2026,munko_multiple_2025,munko_general_2025,zhang2011statistical, zhang_two-sample_2010}. 
The supremum-type statistic is given by
\begin{align}
	S_n(\mathbf{A}, \mathbf{x}) &=  \sup_{t \in \mathcal{T}}T_n(\mathbf{A},\mathbf{x}, t). \label{eq:T_max}
\end{align}
For the integral-type statistic, we additionally assume $\mathcal T=[b,c]\subset\mathbb R,\, b<c,$ to be a compact interval and define
\begin{align}
	I_n(\mathbf{A}, \mathbf{x}) &=   \int_{\mathcal{T}} T_n(\mathbf{A},\mathbf{x},t)\,dt. \label{eq:T_int}
\end{align}
Different to approaches based upon pointwise $t$-type statistics \parencite[e.g.][]{smaga_note_2020, baumeister_resampling-based_2026}, our quadratic-form-based approach directly incorporates the covariance structure of the two different estimators.
The following Theorem states the asymptotic behaviour of both global test statistics under the null hypothesis $\mathcal{H}_0$ and under the alternative $\mathcal{H}_1$ in Equation~(\ref{eq:hypothesis}).

\begin{satz}\label{thm:distribution_test_statistics}
	Let the Assumptions hold.
	\begin{enumerate}
		\item Under $\mathcal{H}_0$, the test statistics $I_n(\mathbf{A},\mathbf{x})$ and $S_n(\mathbf{A},\mathbf{x})$ converge in distribution:
		\begin{align*}
			I_n(\mathbf{A},\mathbf{x})&\stackrel{d}{\longrightarrow}\int_{\mathcal{T}}\mathbf{Z}(t)^\top\,\Xi\left(\mathbf{A},\boldsymbol{\Sigma},t\right)\mathbf{Z}(t)dt,\\
			S_n(\mathbf{A},\mathbf{x})&\stackrel{d}{\longrightarrow}\sup_{t\in\mathcal{T}}\left\{\mathbf{Z}(t)^\top\,\Xi\left(\mathbf{A},\boldsymbol{\Sigma},t\right)\mathbf{Z}(t)\right\},
		\end{align*}
		where $\{\mathbf{Z}(t):\,t\in\mathcal{T}\}$ is the Gaussian process from Proposition~\ref{thm:Aclt}.
		\item Under $\mathcal{H}_1$, $I_n(\mathbf{A},\mathbf{x})$ and $S_n(\mathbf{A},\mathbf{x})$ converge in probability to $\infty$.
	\end{enumerate}
\end{satz}
The asymptotic distributions in Theorem~\ref{thm:distribution_test_statistics} depend on the unknown $\boldsymbol{\Sigma}$, which prevents the direct calculation of critical values. 
As a solution, we approximate the null distribution by parametric bootstrapping.

\subsection{Parametric bootstrap procedure}\label{sec:resampling}
We use the asymptotic model-based parametric bootstrap, which has been successfully applied in the matched-pairs setting \parencite{amro_asymptotic-based_2021} and in various functional testing problems \parencite[e.g.][]{baumeister_resampling-based_2026, cuevas2004anova, martinez2011repeated, martinez2012general, munko_general_2025, munko_multiple_2025, smaga_note_2020}. 
For the concrete implementation we generate resamples from a centred Gaussian process with an estimated covariance function
\begin{align}
	\label{eq:bootstrap_data}
	\mathbf{x}_{j,\mathcal{P}}\sim \GP_2 \left(\mathbf{0},\boldsymbol{\hat{\Gamma}}^{(c)}\right),\,j\in\{1,\dots,n\},
\end{align}
where $\boldsymbol{\hat{\Gamma}}^{(c)}$ is as in Equation~(\ref{eq:cempcovariance}). 
To also obtain the observed missingness structure in the resample, 
we only keep the first $n_1$ complete bootstrap pairs and mark the second component as missing for the remaining $n_2$ bootstrap observations.
This results in the following bootstrap sample
\begin{align*}%\label{eq:bootstrap_missingdata}
	\mathbf{x}_{\mathcal{P}}
	=\underbrace{
		\begin{pmatrix}
			x_{11,\mathcal{P}}^{(c)}(t)\\
			x_{21,\mathcal{P}}^{(c)}(t)
		\end{pmatrix},
		\dots,
		\begin{pmatrix}
			x_{1n_1,\mathcal{P}}^{(c)}(t)\\
			x_{2n_1,\mathcal{P}}^{(c)}(t)
		\end{pmatrix} 
	}_{ \mathbf{x}^{(c)}_{\mathcal{P}}}
	,
	\underbrace{
		\begin{pmatrix}
			x_{11,\mathcal{P}}^{(m)}(t)\\
			---
		\end{pmatrix},
		\dots,
		\begin{pmatrix}
			x_{1n_2,\mathcal{P}}^{(m)}(t)\\
			---
		\end{pmatrix}
	}_ {\mathbf{x}^{(m)}_{\mathcal{P}}}.
\end{align*}
From this we compute the pointwise statistic
\begin{align}\label{eq:pw_bootstrap_teststatistic}
	T^{\mathcal{P}}_n(\mathbf{A},\mathbf{x}_{\mathcal{P}},t)=n\left(\mathbf{A}\boldsymbol{\hat\mu}^{\mathcal{P}}(t)\right)^\top\Xi\left(\mathbf{A},\boldsymbol{\hat\Sigma}^{\mathcal{P}},t\right)\mathbf{A}\boldsymbol{\hat\mu}^\mathcal{P}(t),\,t\in\mathcal{T},
\end{align}
which is the bootstrap version of Equation~(\ref{eq:pw_teststatistic}). 
From this we obtain the bootstrap counterparts of the global test statistics by
\begin{align}
	I^{\mathcal{P}}_n(\mathbf{A}, \mathbf{x}_{\mathcal{P}}) &=   \int_{\mathcal{T}} T^{\mathcal{P}}_n(\mathbf{A},\mathbf{x}_{\mathcal{P}},t) dt, \label{eq:T_int_bootstrap} \\
	S^{\mathcal{P}}_n(\mathbf{A}, \mathbf{x}_{\mathcal{P}}) &=  \sup_{t \in \mathcal{T}}T^{\mathcal{P}}_n(\mathbf{A},\mathbf{x}_{\mathcal{P}}, t). \label{eq:T_max_bootstrap}
\end{align}
The following theorem establishes the asymptotic validity of the bootstrap procedure.
\begin{satz}
	\label{thm:distribution_boot_test_statistic}
	Under the Assumptions, the conditional distributions of the global resampling test statistics approximate the null limiting distributions of the corresponding global test statistics $I_n(\mathbf{A},\mathbf{x})$ respective $S_n(\mathbf{A},\mathbf{x})$ under the null hypothesis and the alternative, i.e., we have under $\mathcal{H}_0$ as well as under $\mathcal{H}_1$:
	\begin{align}
		\sup_{w\in\br}&\left|\prob\left(I_n^\mathcal{P}(\mathbf{A},\mathbf{x}_{\mathcal{P}})\le w|\mathbf{x}\right)-\prob\left(\int_{\mathcal{T}}\mathbf{Z}(t)^\top\,\Xi\left(\mathbf{A},\boldsymbol{\Sigma},t\right)\mathbf{Z}(t)dt\le w\right)\right|\stackrel{p}{\longrightarrow}0,\\
		\sup_{w\in\br}&\left|\prob\left(S_n^\mathcal{P}(\mathbf{A},\mathbf{x}_{\mathcal{P}})\le w|\mathbf{x}\right)-\prob\left(\sup_{t\in\mathcal{T}}\left\{\mathbf{Z}(t)^\top\,\Xi\left(\mathbf{A},\boldsymbol{\Sigma},t\right)\mathbf{Z}(t)\right\}\le w\right)\right|\stackrel{p}{\longrightarrow}0.
	\end{align}
\end{satz}

From this result, we are able to define resampling tests:
\begin{align}\label{eq:boot_tests}
	\varphi_n^{\mathcal{P},I}=\mathds{1}\left\{I_n(\mathbf{A},\mathbf{x})>q^{\mathcal{P},I}_{1-\alpha}(\mathbf{x})\right\};\quad \varphi_n^{\mathcal{P},S}=\mathds{1}\left\{S_n(\mathbf{A},\mathbf{x})>q^{\mathcal{P},S}_{1-\alpha}(\mathbf{x})\right\},
\end{align}
where $q^{\mathcal{P},I}_{1-\alpha}(\mathbf{x})$ and $q^{\mathcal{P},S}_{1-\alpha}(\mathbf{x})$, are the empirical $(1-\alpha)$-quantiles of the conditional distribution functions $w\mapsto \prob(I_n^\mathcal{P}(\mathbf{A},\mathbf{x}_{\mathcal{P}})\le w|\mathbf{x})$ respective $w\mapsto \prob(S_n^\mathcal{P}(\mathbf{A},\mathbf{x}_{\mathcal{P}})\le w|\mathbf{x})$.
It follows from \textcite{janssen_how_2003} that $\varphi_n^{\mathcal{P},\ell}$, $\ell\in\{I,S\}$, are asymptotically exact and consistent.

In practice, the test decisions are realized via Monte-Carlo: Based on $B$ bootstrap replications, the estimated $p$-values are obtained as
\begin{align}\label{eq:boot_pvalues}
	p^{\mathcal{P},\ell} = \frac{1}{B+1}\left(1+\sum_{b=1}^{B} \mathds{1}\left\{\ell_{n,b}^\mathcal{P}(\mathbf{A},\mathbf{x})>\ell_n(\mathbf{A},\mathbf{x})\right\}\right).	
\end{align}
Compare \textcite[][Equation~(17.7)]{lehmann_testing_2022} for the calculation. 
The computation of the $p$-values is summarized in Algorithm~\ref{al:boot_pvalues}.
\begin{algorithm}
	\caption{Calculation of  bootstrap $p$-values.}
\label{al:boot_pvalues}
\begin{algorithmic}[1]
	\State \textbf{Input:} Observed functional data $\mathbf{x}_1(t),\dots,\mathbf{x}_n(t)$,
	with missing values in the second arm, significance level $\alpha$, number of bootstrap samples $B$, type of statistic $\ell\in\{I,S\}$.
	\State Calculate the observed test statistic $\ell_n(\mathbf{A},\mathbf{x})$ and  $\hat{\boldsymbol{\Gamma}}^{(c)}$.
	\For{$b= 1,\dots,B$}
	\State \textbf{Parametric bootstrap:}
	\State Generate $n=n_1+n_2$ independent
	Gaussian processes 
	$\mathbf{x}_{j,\mathcal{P}}\sim \GP_2 \left(\mathbf{0},\boldsymbol{\hat{\Gamma}}^{(c)}\right),$ $j\in\{1,\dots,n\}$ as in~(\ref{eq:bootstrap_data}).
	\State \textbf{Missingness:}
	\State Keep both components for the first $n_1$ observations and only the first component for the remaining $n_2$ observations.         
	\State Compute the bootstrap test statistic $\ell_{n,b}^\mathcal{P}(\mathbf{A},\mathbf{x})$ based on $\mathbf{x}^\mathcal{P}$ as in (\ref{eq:T_int_bootstrap}) or (\ref{eq:T_max_bootstrap}). 
	\EndFor
	\State Calculate the bootstrap $p$-value $p^{\mathcal{P},\ell} = 1/(B+1)(1+\sum_{b=1}^{B} \mathds{1}\{\ell_{n,b}^\mathcal{P}(\mathbf{A},\mathbf{x})>\ell_n(\mathbf{A},\mathbf{x})\})$.
	\State \textbf{Output:} Bootstrap $p$-value $p^{\mathcal{P},\ell}$ and test decision $\mathds{1}\{p^{\mathcal{P},\ell}\leq\alpha\}$
\end{algorithmic}
\end{algorithm}

\section{Simulation Studies}
\label{sec:simulation}
In this section, we conduct a Monte Carlo simulation study to investigate the properties of the proposed tests. 
We thereby investigate their ability to maintain the nominal type I error level and study their empirical power under different mean functions; both for different covariance structures, and missing rates. 
The proposed parametric bootstrap tests are denoted by $I_W$, $I_A$, $I_M$, $S_W$, $S_A$, and $S_M$ corresponding to the integral and supremum version of the Wald-type, \textsc{anova}-type, and modified \textsc{anova}-type statistics.
As benchmarks, we consider the integral and supremum type permutation tests of \textcite{smaga_note_2020}, denoted by $I_{Pc}$ and $S_{Pc}$, respectively. 
As these procedures require fully observed functional pairs, we applied them to the complete pairs only. 
We thereby focus on the permutation based approaches, as they showed the best performance in the simulation study of \textcite{smaga_note_2020}. 
In particular, they performed better than the corresponding bootstrap and asymptotic-based supremum and integral type $t$-tests and the approach of \textcite{Kong_et_al_2026}. 
As additional competitor, we also considered the bootstrap test by \textcite{crainiceanu2012bootstrap}; applied to complete pairs. 
Moreover, we investigated this procedure and the permutation test of \textcite{smaga_note_2020}  combined with the imputation method of \textcite{jang_bayesian_2021}. 
As these three additional approaches exhibited a liberal behaviour in the considered small- to moderate-sample size settings, we only report the results in \texttt{main\_supp.pdf} of the Supplementary Material \textcite{baumeister_supplementary_2026-1}.
The raw data can be found in the file \texttt{simulation\_results\_imputation.csv}.

\subsection{Simulation set-up}
We generate $n=n_1+n_2$ functional observations according to Model~\eqref{eq:data} and subsequently remove $n_2$ functions from the second arm under an \textsc{mcar} mechanism to obtain the data structure \eqref{eq:missingdata}.  
The total sample size is fixed at $n=n_1+n_2=30$, but the number of complete and incomplete pairs differs to investigate the missingness effect.
We consider $\mathbf{n}=(n_1, n_2) \in \{(21,9), (18,12), (15,15), (12,18), (9,21), (6,24)\}$. 
Therefore, the proportion of incomplete observations increases from $30\%$ to $80\%$.
Then, functional data is generated from the model
\begin{align}\label{eq:simu_model}
x_{ij}(t)&= \eta_i(t) + e_{ij}(t),\quad t \in [0,1],\, i \in \{1,2\},\, j\in\{1,\dots,n\}.
\end{align}
All functions are generated on the grid of time points $t_r=r/\iota, r\in\{0,\dots,\iota-1\}$ with $\iota=\{25,50\}$ grid points.

\paragraph{Mean functions.}
We consider six models for the mean functions $\eta_1$ and $\eta_2$:
\begin{align*}
\text{Model } M1: &\ \eta_1(t)=\eta_2(t)=\left\{\sin(2\pi t^2)\right\}^{5},\\
\text{Model } M2: &\ \eta_1(t)\; \text{as in}\; M1,\;\text{and}\; \eta_2(t)=\left\{\sin(2\pi t^2)\right\}^{3},\\
\text{Model } M3: &\ \eta_1(t)=\eta_2(t)=\sqrt{\frac{6t}{\pi}}\exp(-6t),\\
\text{Model } M4: &\ \eta_1(t)\; \text{as in}\; M3,\; \text{and}\; \eta_2(t) = \sqrt{\frac{13t}{2\pi}} \exp\left(-\frac{13t}{2}\right),\\
\text{Model } M5: &\ \eta_1(t)=\eta_2(t)=t^{1/5}(1-t)^5,\\
\text{Model } M6: &\ \eta_1(t)\; \text{as in}\; M5,\; \text{and}\; \eta_2(t) = t^{1/5}(1-t)^{4.7}.
\end{align*}
Models $M1$, $M3$, and $M5$ satisfy the null hypothesis, whereas $M2$, $M4$, and $M6$ represent different alternatives. 
The functions M1-M4 are taken from \textcite{martinez2011repeated} and $M5$-$M6$ from  \textcite{qiu_global_2026}. 

\paragraph{Error processes.}
We consider Gaussian and non-Gaussian error processes. For the Gaussian ones, errors were generated according to
\begin{align*}
e_{1j}(t)=\sigma B_j(t),\ e_{2j}(t)=\rho\sigma B_i(t)+\sigma\sqrt{1-\rho^2}\,B_j^{*}(t),
\end{align*}
where $B_j(t)$ and $B_j^{*}(t)$ are independent standard Brownian bridges and $t \in [0,1]$, $j\in\{1,\dots,n\}$. 
The parameter $\rho$ controls the dependence between the two error processes within each pair and we study 
$$\rho\in\{-0.75,-0.50,-0.25,-0.10,0.10,0.25,0.50,0.75\}.$$ 
Following \textcite{martinez2011repeated} and \textcite{qiu_global_2026}, the scale parameter was set to $\sigma=1/2$ for Models $M1$ and $M2$, and to $\sigma=1/20$ for Models $M3$-$M6$.  
To study the robustness to deviations from Gaussianity, we additionally consider non-Gaussian errors, obtained by exponentiating the Brownian-bridges:
\begin{align}
e_{1j}(t)&=\exp\{\sigma B_j(t)\} - \overline{\exp\{\sigma B_j\}},\label{eq:simu_errornormal}
\\
e_{2i}(t)&= \exp\left\{\rho\sigma B_j(t) +\sigma\sqrt{1-\rho^2}\,B_j^{*}(t) \right\} - \overline{ \exp\left\{ \rho\sigma B_j +\sigma\sqrt{1-\rho^2}\,B_j^{*} \right\} },\label{eq:simu_errorlognormal}
\end{align}
for $t \in [0,1]$, $j\in\{1,\dots,n\}$. 
The bars denote the arithmetic mean over the time grid for each curve. 
This construction produces non-Gaussian and right-skewed functional errors, while preserving the dependence induced by the underlying processes $\{B_j(t): t\in[0,1]\}$, $j\in\{1,\dots,n\}$.

For each combination of mean model, error distribution, grid size, sample sizes, and dependence parameter, we perform $1000$ independent Monte Carlo repetitions.
In each repetition, critical values are obtained using $1000$ bootstrap or permutation samples. 
The nominal significance level was set to $\alpha=0.05$. 
The simulations are implemented in \textsf{R} \parencite[][]{R_Core_Team_2026}, and the corresponding code is given in the Supplementary Material \textcite{baumeister_supplementary_2026-1}.

\subsection{Simulation results}
The complete simulation results are given in the Supplementary Material \parencite[][\texttt{simulation\_results.csv}]{baumeister_supplementary_2026-1}. 
Here, we focus on the main findings using summarizing boxplots. 
Additional figures can be found in the \texttt{main\_supp.pdf} in the Supplementary Material \textcite{baumeister_supplementary_2026-1}.

\paragraph{Type I error control}
\begin{figure}[!t]
\centering
\includegraphics[width=0.8\textwidth]{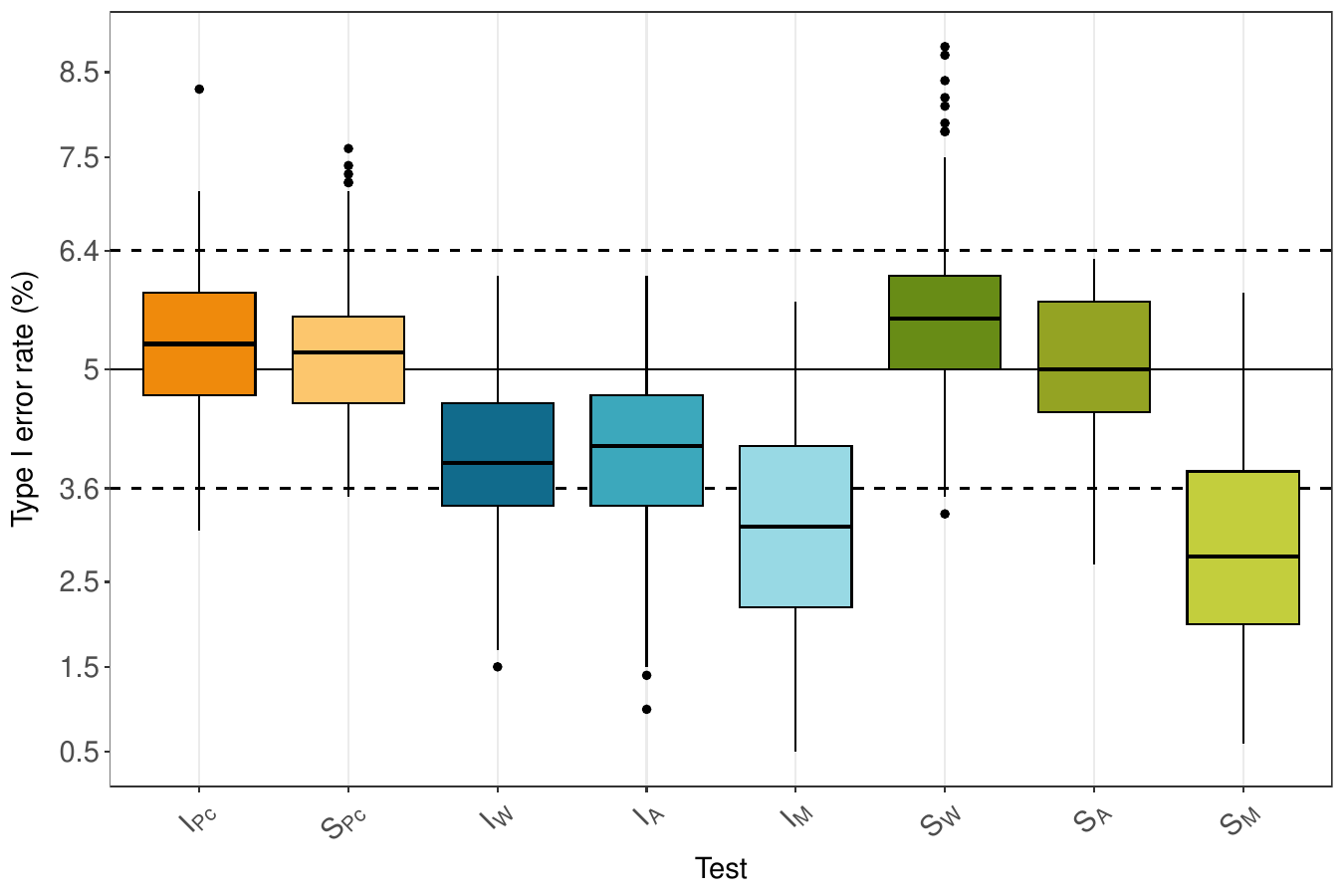}
\caption{Boxplots for the empirical sizes (as percentages) of all tests obtained in Model $M1$. The solid line denotes the significance level $\alpha=5\%$, and the dashed line the limits of the $95\%$ binomial confidence intervals $[3.6\%,6.4\%]$ for 1000 simulation runs \parencite[cf.][]{DuchesneFrancq2015}.}
\label{fig_1}
\end{figure}
We consider the empirical type I error rates aggregated over all simulation settings (Figure~\ref{fig_1} and Figures~1-2 in the Supplementary Material). 
In all figures, the solid horizontal line indicates the nominal level of $5\%$, while the dashed lines show the binomial confidence interval from $3.6\%$ to $6.4\%$ used to assess empirical sizes \parencite[][]{DuchesneFrancq2015}. 
The complete-case permutation tests of \cite{smaga_note_2020}, $I_{Pc}$ and $S_{Pc}$, keep the prescribed nominal level quite well and are only a bit too liberal in somce settings.  
Among the newly proposed procedures, all three integral tests ($I_W$, $I_A$, $I_M$) as well as the modified supremum test $S_M$ control the type I error rate across all considered scenarios with empirical sizes always below the upper binomial level of $6.4\%$. 
However, they are also quite conservative, especially $I_M$ and $S_M$ which exhibit
empirical sizes below $3.6\%$ in several settings. In contrast, the supremum Wald-type test $S_W$ can sometimes be a bit liberal. 
Overall, the \textsc{anova}-type supremum test $S_A$ provides the most stable type I error control. 

\begin{figure}[!t]
\centering
\includegraphics[width=0.99\textwidth]{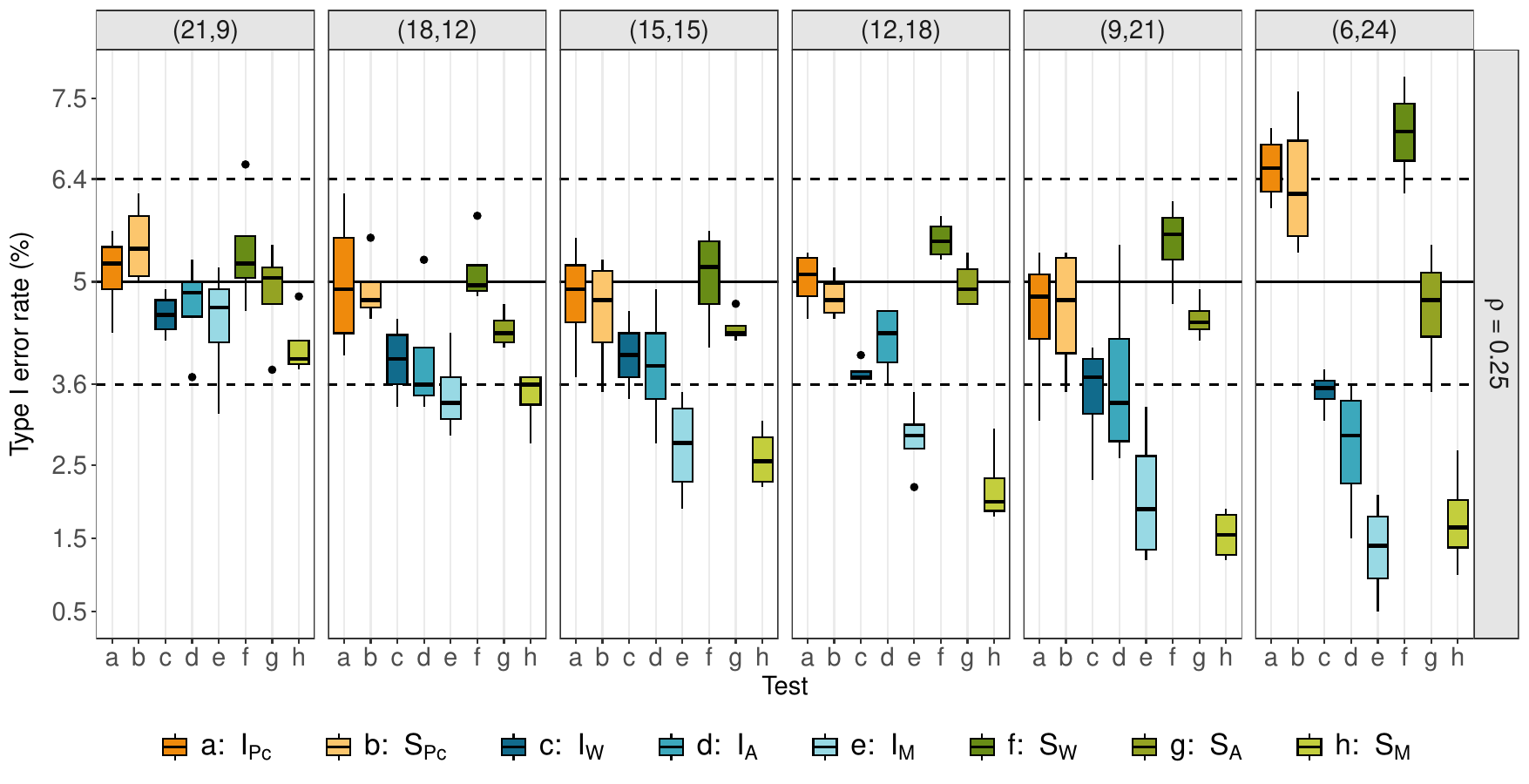}
\caption{Boxplots for the empirical sizes (as percentages) of all tests obtained in Model $M1$ for different sample sizes and $\rho=0.25$. The solid line denotes the significance level $\alpha=5\%$, and the dashed line the limits of the $95\%$ binomial confidence intervals $[3.6\%,6.4\%]$ for 1000 simulation runs \parencite[cf.][]{DuchesneFrancq2015}.}
\label{fig_6}
\end{figure}

To get further insides, we differentiated the results by certain characteristics. 
Figure~\ref{fig_6} gives the empirical sizes for $\rho=0.25$, separated according to the numbers of complete and incomplete observations. 
Analogous figures for the remaining values of $\rho$ are presented in the Supplementary Material \parencite[][\texttt{main\_supp.pdf}, Figures~3-8]{baumeister_supplementary_2026-1}. 
Again, the tests $I_{Pc}$ and $S_{Pc}$ remain quite stable as they use only the complete pairs. 
Nevertheless, their variability increases with a decreasing number of complete pairs, ultimately leading to a liberal behaviour in the setting with the highest missingness ($\mathbf{n}=(6,24)$). 
Outside this missing configuration, liberal results are uncommon for these procedures and do not show a systematic relationship with the mean model.
The properties of the six proposed bootstrap tests varies more with a decreasing number of complete observations. 
In fact, all integral tests and the $S_M$ test become increasingly conservative when the proportion of incomplete observations grows. 
This effect is particularly strong for $I_M$ and $S_M$, while $I_W$ and $I_A$ show a similar but less pronounced decrease. 
As a consequence, these procedures may have reduced power when missingness is high.  
In contrast, $S_W$ becomes liberal with increasing missingness while $S_A$ is again the most stable approach.

The results for the other mean models under the null hypothesis ($M3$ and $M5$) are qualitatively similar, compare Figure 1 and 2 in \texttt{main\_supp.pdf} of \textcite{baumeister_supplementary_2026-1}. 
Furthermore, increasing the number of grid points from 25 to 50 did not produce a systematic improvement or deterioration.
Differentiating between the error distributions in Equations~(\ref{eq:simu_errornormal}) and~(\ref{eq:simu_errorlognormal}), however, lead to a higher variability. 
Under non-Gaussian errors, $I_W$, $I_A$, $I_M$, and $S_M$ are even more conservative while $S_W$ becomes more liberal. 
Liberal results for $S_W$ also become more frequent under positive correlations and are usually strongest for large positive values of $\rho$. 
Overall, $S_A$ offers the most stable type I error control among the proposed tests. 
The integral tests $I_W$ and $I_A$ are also competitive, but can become conservative in case of high missingness.

\paragraph{Power}
\begin{figure}[!t]
\centering
\includegraphics[width=0.99\textwidth]{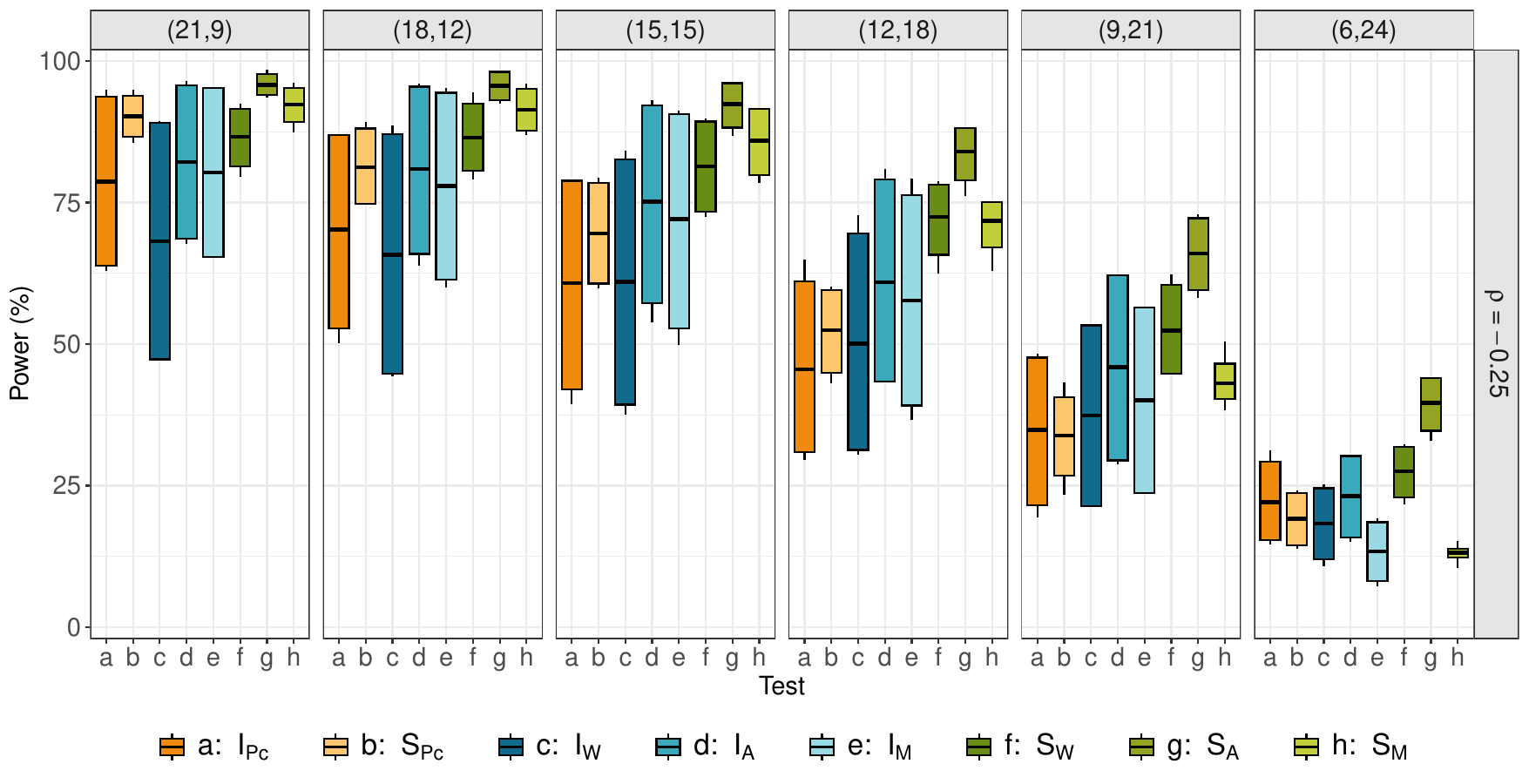}
\caption{Boxplots for the empirical powers (as percentages) of all tests obtained in Model $M2$ for different sample sizes and dependence parameter $\rho=-0.25$.}
\label{fig_11}
\end{figure}
The power results are summarized in Figures~\ref{fig_11}-\ref{fig_13} and Figures~9-11 in the document \texttt{main\_\-supp.pdf} of the supplementary material \textcite{baumeister_supplementary_2026-1}.
Figure~\ref{fig_11} shows the empirical power for Model $M2$ for different sample sizes in case of $\rho=-0.25$. 
Analogous plots for other values of $\rho$ are given in Figure~9 in the Supplementary Material. 
In all plots, power generally decreases with increasing missingness. 
This is expected, as fewer complete pairs provide less information about the second component and the within-pair dependence. 
Moreover, the competing complete-case permutation tests $I_{Pc}$ and $S_{Pc}$ lose considerable power as they discard all incomplete observations. 
For the newly proposed tests this effect should be smaller as they still use all available information, which can be readily seen for the supremum-type tests. 
In fact, for Model $M2$, the supremum-based tests are usually more powerful than their integral counterparts. 
Among the proposed procedures, $S_A$ generally exhibits the highest power.
This is surprising as it also showed the most stable type I error control; especially compared to the sometimes rather liberal $S_W$.
The power of $S_M$ is also competitive up to $\mathbf{n}=(15,15)$ but decreases rapidly with increasing missingness which is in line with its conservative behaviour under the null.

\begin{figure}[!t]
\centering
\includegraphics[width=\textwidth]{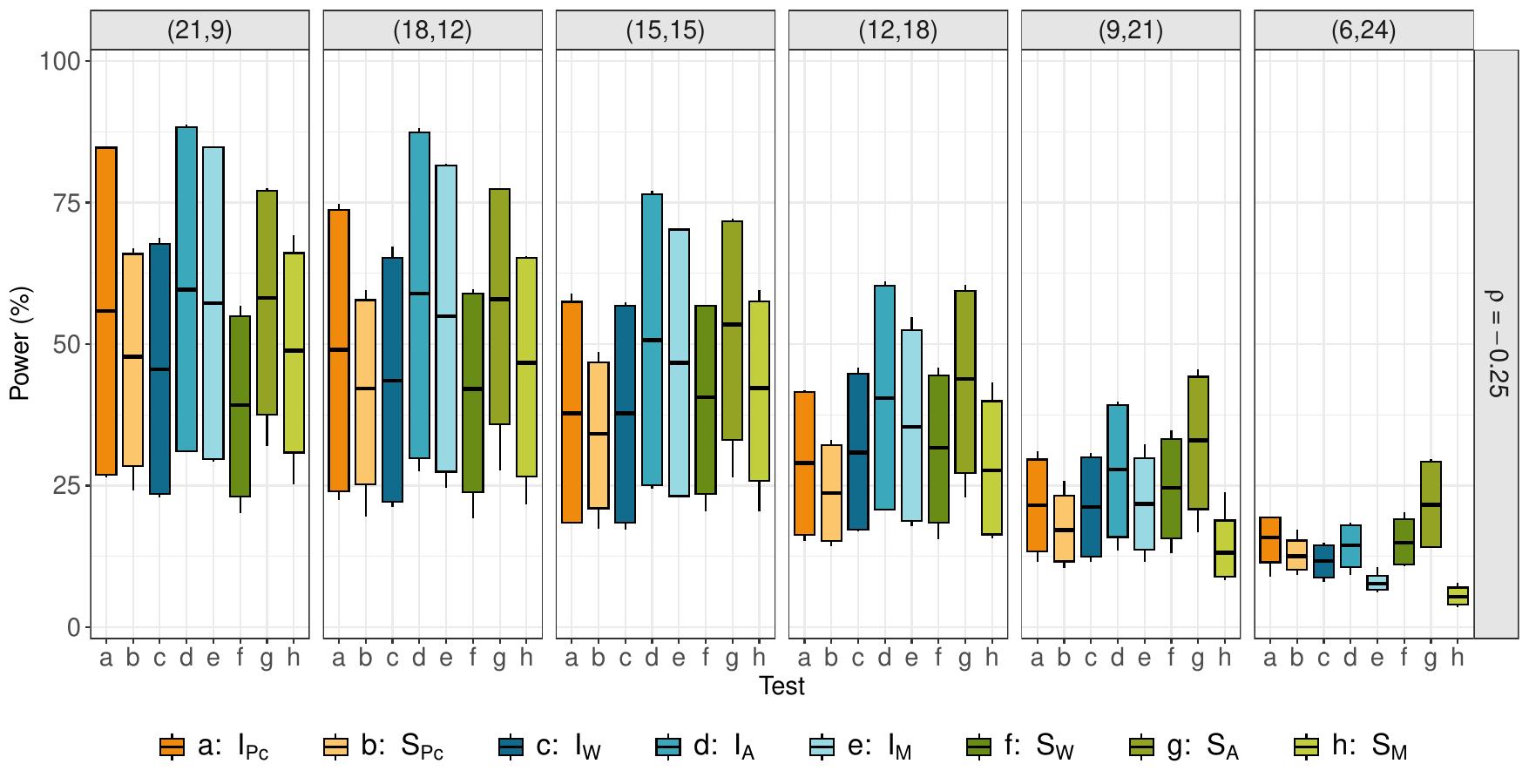}
\caption{Boxplots for the empirical powers (as percentages) of all tests obtained in Model $M4$ for different sample sizes and dependence parameter $\rho=-0.25$.}
\label{fig_12}
\end{figure}
In Figure~\ref{fig_12}, the corresponding results for Model $M4$ are presented (see also Figure~10 in the Supplementary Material for other values of $\rho$ than $-0.25)$) and it is observable that the overall power is lower than for Model $M2$. 
Here, no clear differences between the integral and supremum versions of the tests are visible. 
The supremum \textsc{anova}-type test $S_A$ remains powerful, but the integral versions $I_W$ and $I_A$ are similar strong in the two cases with at least 18 complete pairs. 
Moreover, the relative performance for Model $M4$ depends strongly on the data generating process. 
Under Gaussian error processes, $S_A$ is generally the most powerful test. 
In the non-Gaussian setting, however, $I_A$ becomes highly competitive and often outperforms $S_A$. 
Therefore, $I_A$ is an attractive procedure in these settings, especially as it also controls the type I error rate in all considered scenarios. 
Once again, the modified tests $I_M$ and $S_M$ are usually less powerful, especially when the proportion of incomplete observations is large which is again consistent with their conservative behaviour under the null.

\begin{figure}[!t]
\centering
\includegraphics[width=0.99\textwidth]{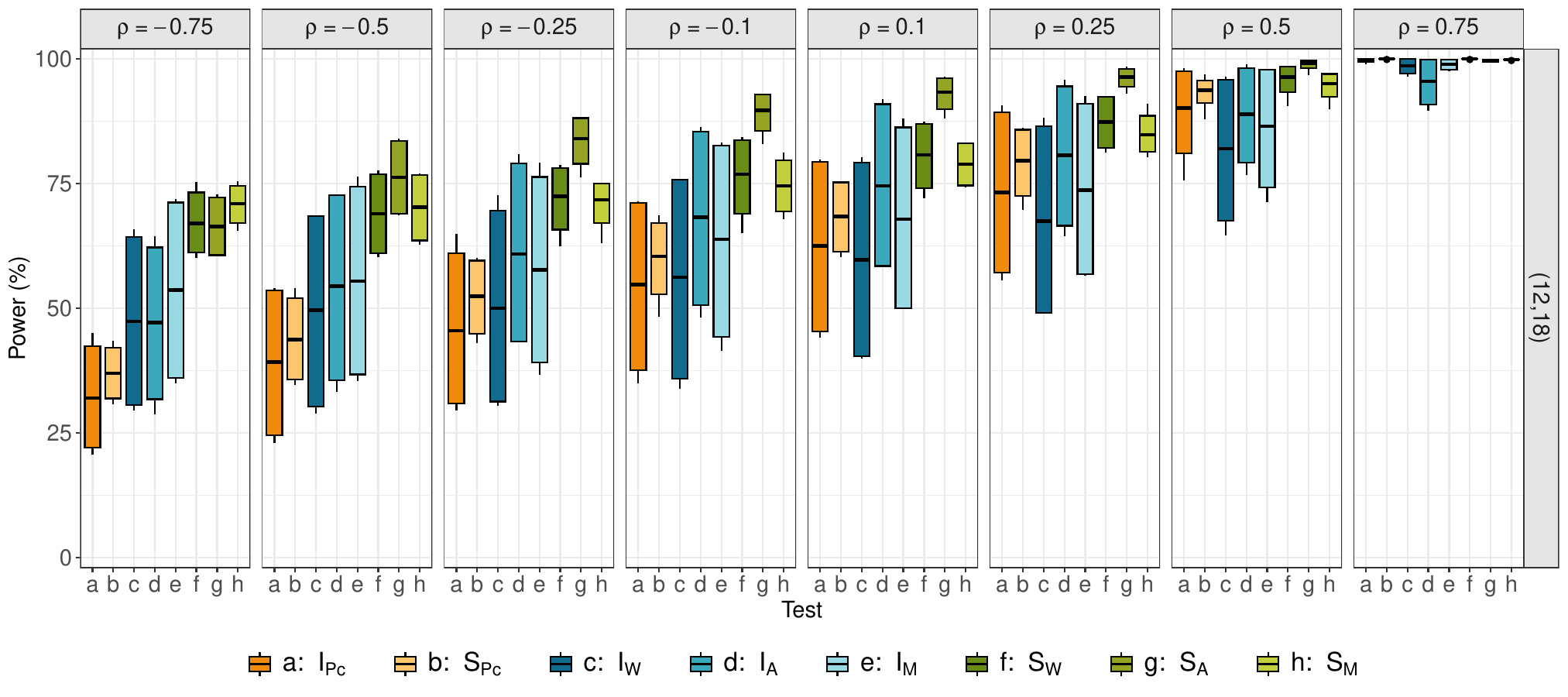}
\caption{Boxplots for the empirical powers (as percentages) of all tests obtained in Model $M4$ for different dependence parameters $\rho$ and $\mathbf{n}=(12,18)$.}
\label{fig_13}
\end{figure}
Figure~\ref{fig_13} displays the empirical powers for a fixed sample-size configuration of $\mathbf{n}=(12,18)$ and varying dependence parameters $\rho$. 
Here, power generally increases as $\rho$ changes from strongly negative to strongly positive. 
As a result, all tests have relatively low power for negative values of $\rho$, whereas their powers become very high when $\rho$ is strongly positive. 
For $\rho=0.75$, most tests have empirical power close to $100\%$. 
This behaviour is expected as positive dependence reduces the variability of the complete-pairs (since for two random variables X and Y: $Var(X-Y) = Var(X) + Var(Y) - 2 Cov(X,Y)$), leading to a better signal-to-noise ratio. 
Regarding power, $S_A$ remains one of the strongest procedures across all dependence settings. 
The test $I_A$ is also competitive, particularly under non-Gaussian errors. 
The competing permutation tests $I_{Pc}$ and $S_{Pc}$ are worse in cases with $\rho\leq -0.25$ but perform well in some settings with strong positive dependence ($\rho\geq 0.5$).  
The simulation results for Model $M6$ lead to similar conclusions (see Figure~11 in the Supplementary Material). 
Under Gaussian errors, $S_A$ is generally the most powerful procedure while $I_A$ and $S_A$ are comparable in case of non-Gaussian error processes.

\paragraph{Concluding Recommendation.}
Overall, in our simulations, $S_A$ provides the best compromise between good empirical power and accurate type I error control. 
The test $I_A$ is especially competitive for non-Gaussian processes and for alternatives that are better detected by integral aggregation.

\section{Illustrative Data Analysis of DTI Profiles}
\label{sec:real_data_example}
We illustrate the proposed methods using Diffusion Tensor Imaging (\textsc{dti}) data which was collected at Johns Hopkins University and the Kennedy Krieger Institute. 
The data is available in the  \textsf{R} package \texttt{refund} \parencite[][]{Goldsmith_et_al_2026}. 
\textsc{dti} is a magnetic resonance imaging technique that provides various measures of water diffusivity along white matter tracts in the brain. 
It is useful for studying diseases that affect brain white matter, such as multiple sclerosis (\textsc{ms}). 
To investigate the impact of neurodegeneration on disability, \textsc{dti} brain scans were collected from \textsc{ms} patients during multiple visits. 
Based on these scans, fractional anisotropy (\textsc{fa}) tract profiles were obtained for the corpus callosum (\textsc{cca}). 
Each patient's \textsc{fa} profile is observed at 93 design points at every visit and is therefore considered as functional observation defined on the interval $[1,93]$.

The number of visits varies from two to eight, with fewer patients examined in subsequent visits. 
For illustrative purposes, we focus on the second and third visits, considering patients who participated in both visits as well as from patients who participated only in the second visit.
Since observations from the two visits are paired for each patient, while additional observations are available only from visit two, the data has a matched-pairs structure with missingness in a single arm. 
We can therefore apply the methods considered in the simulation study to test the null hypothesis of equal mean functions of the two visits: the method by \textcite{smaga_note_2020} for paired functional data based on the complete pairs only and the proposed methods that additionally incorporate the incomplete observations.

\begin{figure}[!t]
\centering
\includegraphics[width=0.99\textwidth]{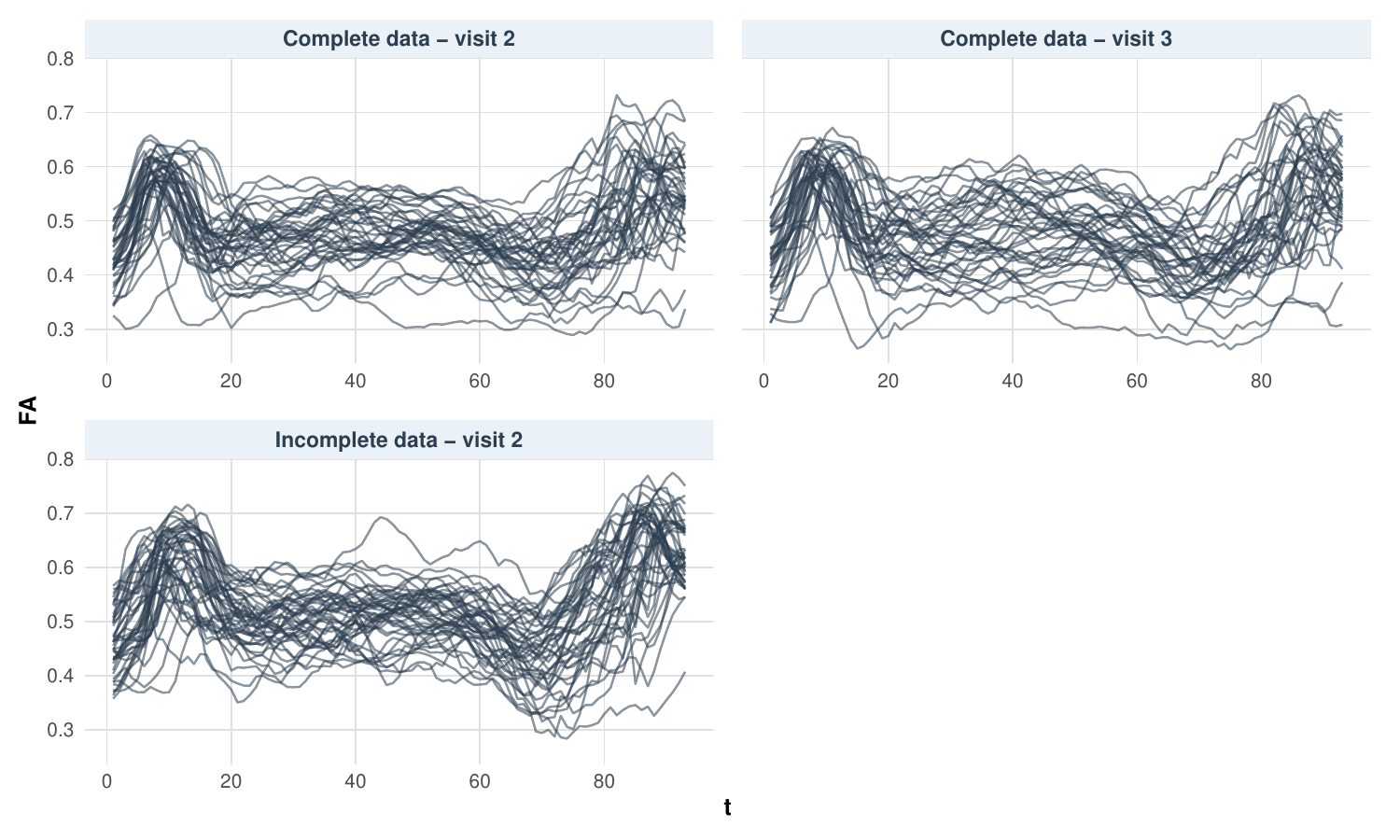}
\caption{Individual \textsc{fa} profiles. The upper panels show the complete observations from visits two and three, while the lower panel shows the additional observations from patients who participated only in the second visit.}
\label{fig_18}
\end{figure}

\begin{figure}[!t]
\centering
\includegraphics[width=0.99\textwidth]{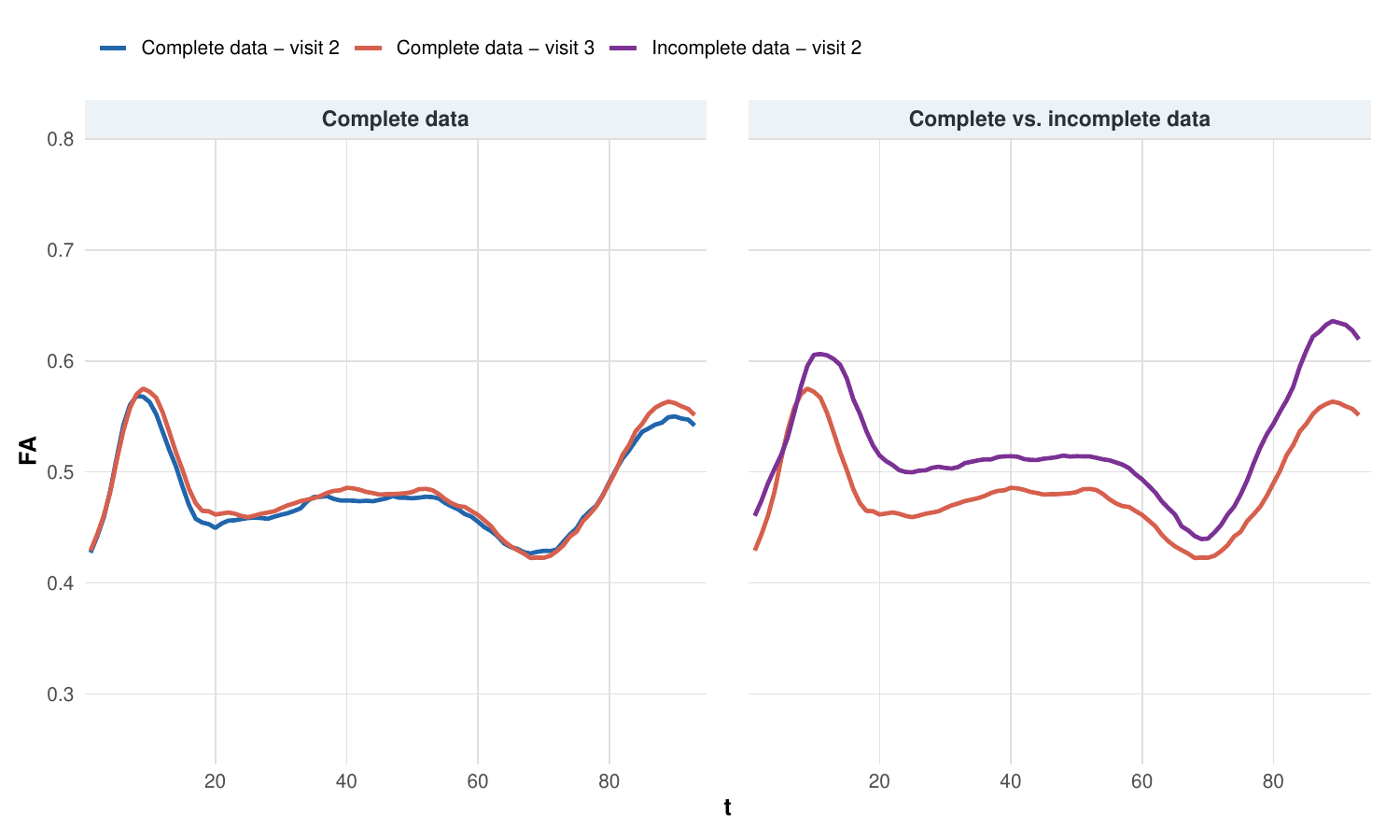}
\caption{Sample mean \textsc{fa} profiles. The left panel compares the complete observations from visits two and three. The right panel compares the complete observations from visit three with the incomplete observations from visit two.}
\label{fig_19}
\end{figure}

For visits two and three, the data set contains $n_1=54$ complete pairs and $n_2=44$ additional functional observations from visit two, i.e. we have $n=98$ patients. 
When all complete pairs were included, all tests rejected the null hypothesis. 
To obtain a more informative comparison of the procedures, we restricted our subsequent illustrative analysis to the first $n_1=40$ complete pairs and the $n_2=44$ incomplete observations. 
The trajectories in the three resulting samples are presented in Figure~\ref{fig_18}, and the corresponding sample mean functions are displayed in Figure~\ref{fig_19}.

The sample mean functions for the complete observations at visits two and three are quite similar across the entire domain (cf. left panel of Figure~\ref{fig_19}). 
Detecting a difference between the mean functions using only the complete pairs may therefore be difficult. 
In contrast, the right panel of Figure~\ref{fig_19} shows a difference between the mean function for the incomplete observations from visit two and that for the complete observations from visit three. 
In particular, the mean \textsc{fa} values for the complete observations from visit three are generally lower than those for the incomplete observations from visit two. 
This pattern is consistent with longitudinal findings indicating that disease progression in patients with active \textsc{ms} may be accompanied by a reduction in \textsc{fa} in the corpus callosum \parencite[][]{Tian_et_al_2012}. 
However, this interpretation should be treated with caution because the complete and incomplete observations come from different groups of patients, and the observed difference cannot be attributed to disease progression alone.

The results of the tests are presented in the second column of Table~\ref{tab_1}. 
The complete-pairs tests $I_{Pc}$ and $S_{Pc}$ yielded $p$-values of $0.090$ and $0.061$, respectively. 
Therefore, neither test rejected the null hypothesis of equality of the mean functions at the significance level of 0.05. 
These findings are consistent with the observation that the two mean functions are quite similar for the complete pairs. 
In contrast, all procedures incorporating the incomplete observations rejected the null hypothesis with $p$-values between $0.001$ and $0.002.$. 
This substantial difference between the two groups of tests illustrates the potential effect of using the additional incomplete observations for inference. 
At the same time, the differences between the complete and incomplete observations may also reflect systematic differences between patient groups as discussed above. 

\begin{table} \centering \setlength{\tabcolsep}{6pt} \renewcommand{\arraystretch}{1.15} 
\caption{$P$-values and empirical sizes and powers for the \textsc{dti} data set} 
\begin{tabular}{lcrrr@{\hspace{20pt}}rrr} %\begin{tabular}{lcrrrrrrr}
	\toprule %{ l S[table-format=1.3] *{6}{S[table-format=2.2]} } \toprule
	\multicolumn{1}{c}{\multirow{2}{*}{Test}} & \multicolumn{1}{c}{\multirow{2}{*}{$P$-value}} & \multicolumn{3}{c}{\makebox[0pt][c]{Empirical size (\%)}} & \multicolumn{3}{c}{\makebox[0pt][c]{Empirical power (\%)}} \\ \cmidrule(lr){3-5} \cmidrule(lr){6-8} %\cmidrule(lr){3-5} \cmidrule(lr){7-9} 
	& & $N$ & $t_5$ & $\chi_{10}^2$ & $N$ & $t_5$ & $\chi_{10}^2$ \\ 
	\midrule $I_{Pc}$ & 0.090 & 4.6 & 4.9 & 5.0 & 61.0 & 68.6 & 59.9\\ 
	$S_{Pc}$ & 0.061 & 5.2 & 6.4 & 4.7 & 73.3 & 81.4 & 71.7\\ \addlinespace[2pt] 
	\midrule \addlinespace[2pt] 
	$I_W$ & 0.001 & 5.4 & 4.9 & 3.5 & 100.0 & 99.9 & 100.0 \\ 
	$I_A$ & 0.001 & 5.2 & 4.7 & 4.8 & 98.7 & 97.2 & 97.8 \\ 
	$I_M$ & 0.001 & 5.4 & 4.9 & 3.9 & 99.5 & 99.0 & 99.7 \\ 
	$S_W$ & 0.001 & 5.2 & 4.5 & 4.3 & 100.0 & 99.9 & 100.0 \\ 
	$S_A$ & 0.001 & 5.4 & 5.7 & 5.2 & 99.8 & 100.0 & 99.9 \\ 
	$S_M$ & 0.002 & 4.8 & 4.2 & 2.9 & 99.6 & 99.4 & 99.6 \\ 
	\bottomrule 
\end{tabular} 
\label{tab_1}
\end{table}

To further study the behaviour of the procedures under realistic scenarios, we run additional simulation studies based on the \textsc{dti} data. 
The simulated samples were designed to reflect the structure of the selected \textsc{dti} data set and were generated according to the Model~\eqref{eq:simu_model} in the following way: we simulated $n_1=40$ complete pairs and $n_2=44$ incomplete observations. 
The covariances were estimated directly from the observations, i.e. the covariance function for the (in)complete observations was set equal to the sample covariance function estimated from the (in)complete pairs. 
To evaluate the type I error rate of the tests, all functional observations were generated under a common mean function ($\eta_1=\eta_2$) equal to the sample mean function of the pooled sample of $2n_1+n_2=124$ observations. 
To assess the power, observations in each sample were generated using the sample mean function estimated from the corresponding group in the data set. 
We considered three distributional settings for the error processes $e_{ij}(t)$ in the Model~\eqref{eq:simu_model}. 
Let $\hat{\boldsymbol{\Psi}}$ denote the estimated covariance matrix on the observation grid for either a complete pair or an incomplete observation described above. 
The corresponding error vectors were generated from: (1) a multivariate normal distribution $\mathcal{N}(\mathbf{0},\hat{\boldsymbol{\Psi}})$, (2) a multivariate Student's $t$ distribution with five degrees of freedom and scale matrix $(3/5)\hat{\boldsymbol{\Psi}}$, or (3) as $\hat{\boldsymbol{\Psi}}^{1/2}\mathbf{Z}$, where the entries of $\mathbf{Z}$ were independent standardized $\chi_{10}^2$ random variables, i.e. $(\chi_{10}^2-10)/\sqrt{20}$. 
In all three settings, the error vectors had zero mean and covariance matrix $\hat{\boldsymbol{\Psi}}$.

The simulation results are also reported in Table~\ref{tab_1}. 
Under the null hypothesis, the empirical rejection rates were generally close to the nominal significance level of $5\%$ for all procedures and all three distributions. 
Only under the $\chi_{10}^2$ distribution, a slight conservativeness was observed for some procedures, particularly for $S_M$, whose empirical size was $2.9\%$. 
Overall, the tests provided satisfactory control of the type I error rate in these settings. 

The power results are consistent with the findings from the illustrative analysis of the original data. 
The complete-pair tests exhibited moderate power, ranging from  $59.9\%$ to $68.6\%$ for $I_{Pc}$ and from $71.7\%$ to $81.4\%$ for $S_{Pc}$. 
The higher power of $S_{Pc}$ is consistent with its smaller $p$-value in the illustrative analysis. 
In contrast, the procedures incorporating incomplete observations had substantially higher power under all three distributions with values ranging from $97.2\%$ to $100\%$. 
Therefore, the observed differences between the procedures in the original data analysis is also present in this data-based simulation study.  

Overall, these illustrative results exemplify the advantage of incorporating incomplete observations into inference procedures for functional matched-pairs.

\section{Conclusions}\label{sec:end}
We have introduced new testing procedures for the equality of mean functions in matched functional pairs with missingness in a single arm. 
Under a \textsc{mcar} mechanism, our approach combines information from complete and incomplete pairs and therefore avoids complete-case analysis and imputation. 
The tests are based on pointwise quadratic form-type statistics that account for the dependence between the involved two estimators of the mean difference. 
Global test statistics are obtained by supremum or integral functionals. 
As their asymptotic null distributions are non-pivotal, parametric bootstrap was used to approximate critical values. 
Using empirical process theory, we have proven the asymptotic validity and consistent of the resulting tests.

To assess the finite sample properties of the new methods, an extensive simulation study was conducted under different data generating processes, varying distribution, mean functions, dependence structure, and missingness.
 We thereby compared their empirical type I error control and power with existing complete-case and imputation approaches.
The results show that type I error control becomes more difficult in cases with increasing missingness, while the power depends highly on the dependency structure of the pairs and the missingness.
Among the considered procedures, the supremum-based \textsc{anova}-type statistic $S_A$ showed the most favourable overall performance, exhibiting the most stable type I error control and a very competitive power behaviour.

Several extensions are of interest for future research. 
In particular, more general repeated measures and multivariate factorial designs with functional observations and missingness remain to be investigated.
Moreover, the parametric bootstrap turned out to be feasible in our setting. 
However, it would be interesting to investigate other resampling procedures, particularly general permutation-type approaches as considered in a special case in \textcite{baumeister_resampling-based_2026}. 
Apart from that, extending the underlying missing mechanism to missing at random \textsc{mar} would be relevant. 
Finally, while the \textsf{R}-code of the data analysis and the simulations is provided in the Supplementary Material \textcite{baumeister_supplementary_2026-1} and can be used for own analyses, we also plan a software implementation in an \textsf{R}-package to facilitate application.

\appendix

\section*{Data Availability Statement}
The data that support the findings of this work, e.g., all simulation results, simulation scripts and the scripts for the data analysis are openly available in TUDOdata at \url{https://doi.org/10.17877/TUDODATA-2026-MWTCDBX9}.
The data considered in the data analysis in Section~\ref{sec:real_data_example} is publicly available in the \textsf{R}-package \texttt{refund} \parencite[][]{Goldsmith_et_al_2026}.

\section*{Acknowledgements}
The work of M. Baumeister and M. Pauly were supported by the German Research Foundation project, Grant no. 565106208.
The work of Ł. Smaga was funded by the National Science Centre, Poland under the Weave-UNISONO (grant no. 2025/07/Y/ST6/00001). 
A part of calculations for the simulation studies was made at the Pozna{\^n} Supercomputing and Networking Center (grant no. Grant pl0253-04).
We thank our student assistant Antonia Köhr for her help with programming and plotting in \textsf{R}.

\section*{CRediT Authorship Contribution Statement}
\textbf{Marléne Baumeister:} Conceptualization, Methodology, Formal analysis, Data Curation, Writing - Original Draft, Visualization, Project administration;
\textbf{Markus Pauly:} Conceptualization, Methodology, Writing - Review \& Editing, Supervision, Funding acquisition;
\textbf{\L ukasz Smaga:} Conceptualization, Software, Validation, Formal analysis, Data Curation, Writing - Original Draft, Visualization, Funding acquisition.

\section{Proofs and Mathematical Theory}\label{sec:proofs}
In this section, some aspects of empirical process theory that are useful to understand the mathematical modelling in this paper, and the technical proofs of all stated theorems are presented.

\subsection{Modelling Functional Data with Empirical Processes}\label{sec:eptheory}
We follow the path of \textcite{dette_functional_2020, munko_multiple_2025,baumeister_resampling-based_2026} and handle functional data as realizations of stochastic processes valued in $C^p(\mathcal{T})$ equipped with the norm $\|f\|_\infty=\sup_{t \in \mathcal{T}}|f(t)|$. 
Consider the proof section of \textcite{baumeister_resampling-based_2026} for more details and note that the Condition~(\ref{eq:packingnumber}) in the Assumptions has a connection to Theorem 2.2.4 in \textcite{van_der_vaart_weak_2023}.
For didactic reasons, we state the following central limit theorem which forms the basis of our method and has already been formulated in \textcite[][Proof Section Theorem 7]{baumeister_resampling-based_2026}.
It is a special case of Lemma 5 in \textcite{munko_multiple_2025}.
\begin{satz}\label{thm:clt}
Under \ref{ass:A5}, it holds
\begin{align}\label{eq:clt}
	\left(\sqrt{n}(\boldsymbol{\hat\eta}-\boldsymbol{\eta})(t)\right)_{t\in\mathcal{T}}\stackrel{d}{\longrightarrow}\mathbf{Y}=(Y_1,Y_2)^\top\sim \GP_2(\mathbf{0},\boldsymbol{\Gamma})
\end{align}
for $n\to\infty$ in $C^2(\mathcal{T})$.
\end{satz}

\subsection{Proof of Proposition~\ref{thm:Aclt}}\label{sec:proof_Aclt}
From Theorem~\ref{thm:clt}, it follows with \ref{ass:A5} that
\begin{align*}
\left(\sqrt{n_1}(\boldsymbol{\hat\eta}^{(c)}-\boldsymbol{\eta})(t)\right)_{t\in\mathcal{T}}\stackrel{d}{\longrightarrow}\mathbf{Y}=(Y_1,Y_2)^\top\sim \GP_2(\mathbf{0},\boldsymbol{\Gamma})
\end{align*}
for $n_1\to\infty$ in $C^2(\mathcal{T})$ and 
\begin{align*}
\left(\sqrt{n_2}(\hat\eta_1^{(m)}-\eta_1)(t)\right)_{t\in\mathcal{T}}\stackrel{d}{\longrightarrow}Y_1\sim \GP(0,\gamma_1)
\end{align*}
for $n_2\to\infty$ in $C(\mathcal{T})$. 
Both convergences can be described together regarding the full sample size $n=n_1+n_2$ due to Assumption~\ref{ass:A4}:
\begin{align*}
\left(\sqrt{n}(\boldsymbol{\hat\mu}-\boldsymbol{\mu})(t)\right)_{t\in\mathcal{T}}&=\sqrt{n}\,\begin{pmatrix}
	\sqrt{n_1}^{-1}\sqrt{n_1}\;(\boldsymbol{\hat\eta}^{(c)}-\boldsymbol{\eta})(t)\\
	\sqrt{n_2}^{-1}\sqrt{n_2}\;(\hat\eta_1^{(m)}-\eta_1)(t)
\end{pmatrix}_{t\in\mathcal{T}}
\stackrel{d}{\longrightarrow}
\begin{pmatrix}
	\kappa_1^{-1}\mathbf{Y}\\
	\kappa_2^{-1}Y_1
\end{pmatrix}\sim\GP_3(\mathbf{0},\boldsymbol{\Sigma})
\end{align*}
in $C^3(\mathcal{T})$ with
\begin{align*}
\boldsymbol{\Sigma}(s,t)=\begin{pmatrix}
	\kappa_1^{-1}\boldsymbol{\Gamma}(s,t) & \mathbf{0}_2\\
	\mathbf{0}_2^\top & \kappa_2^{-1}\gamma_1(s,t)
\end{pmatrix}.
\end{align*}
Therefore, the assertion follows from the Continuous Mapping Theorem.

\subsection{Proof of Theorem~\ref{thm:distribution_test_statistics}}\label{sec:proof_distribution_test_statistics}

We first consider the asymptotic behaviour of the different mid parts $\Xi$ in Proposition~\ref{thm:xi_consistent}.
Afterwards, we prove the assertions of Theorem~\ref{thm:distribution_test_statistics}, e.g., the asymptotic behaviour of the test statistics  under the null hypothesis $\mathcal{H}_0$ and the alternative $\mathcal{H}_1$.
Throughout, $\vertiii{\cdot}$ denotes the spectral norm.
\begin{prop}\label{thm:xi_consistent}
Under the Assumptions, the different versions of  $\Xi(\mathbf{A},\boldsymbol{\hat\Sigma},t)$ under study, namely
\begin{align*}
	\Xi_{\textsc{wts}}\left(\mathbf{A},\boldsymbol{\hat\Sigma},t\right)&=\left(\mathbf{A}\boldsymbol{\hat\Sigma}(t,t)\mathbf{A}^\top\right)^+,\\
	\Xi_{\textsc{ats}}\left(\mathbf{A},\boldsymbol{\hat\Sigma},t\right)&=\tr\left(\mathbf{A}\boldsymbol{\hat\Sigma}(t,t)\mathbf{A}^\top\right)^{-1},\\
	\Xi_{\textsc{mats}}\left(\mathbf{A},\boldsymbol{\hat\Sigma},t\right)&=\left(\mathbf{A}\boldsymbol{\hat\Sigma}(t,t)\mathbf{A}^\top\right)_0^+,
\end{align*}
are uniformly consistent in the following sense:
\begin{align}
	\sup_{t \in \mathcal{T}} \vertiii{\left(\mathbf{A}\boldsymbol{\hat\Sigma}(t,t)\mathbf{A}^\top\right)^+ - \left(\mathbf{A}\boldsymbol{\Sigma}(t,t)\mathbf{A}^\top\right)^+} &\stackrel{p}{\longrightarrow} 0, \label{eq:xi_wts_consistent}\\
	\sup_{t \in \mathcal{T}} \vertiii{\tr\left(\mathbf{A}\boldsymbol{\hat\Sigma}(t,t)\mathbf{A}^\top\right)^{-1} - \tr\left(\mathbf{A}\boldsymbol{\Sigma}(t,t)\mathbf{A}^\top\right)^{-1}} &\stackrel{p}{\longrightarrow} 0, \label{eq:xi_ats_consistent}\\
	\sup_{t \in \mathcal{T}} \vertiii{\left(\mathbf{A}\boldsymbol{\hat\Sigma}(t,t)\mathbf{A}^\top\right)_0^+ - \left(\mathbf{A}\boldsymbol{\Sigma}(t,t)\mathbf{A}^\top\right)_0^+} &\stackrel{p}{\longrightarrow} 0. \label{eq:xi_mats_consistent}
\end{align}
\end{prop}
\begin{proof}[Proof of Proposition~\ref{thm:xi_consistent}]
As stated above, because of the weak law of large numbers, $\boldsymbol{\hat\Gamma}^{(c)}(s,t)$ is point wisely consistent for every $s,t\in\mathcal{T}$.
Furthermore, Lemma~6 in \textcite{munko_multiple_2025} ensures that $\boldsymbol{\hat\Gamma}^{(c)}$ is uniformly consistent over $\{(t,t)\in\mathcal{T}^2\,|\,t\in\mathcal{T}\}$ with the Assumptions \ref{ass:A2} and \ref{ass:A5}.
As $\boldsymbol{\hat\Sigma}(t,t)$ is just a combination of parts of $\boldsymbol{\hat\Gamma}^{(c)}(t,t)$ for every $t\in\mathcal{T}$, $\boldsymbol{\hat\Sigma}$ is uniformly consistent over $\{(t,t)\in\mathcal{T}^2\,|\,t\in\mathcal{T}\}$ as well under Assumptions~\ref{ass:A2}, \ref{ass:A4}, and \ref{ass:A5}.
By Assumption~\ref{ass:A3}, $\boldsymbol{\Gamma}(t,t)$ is non-singular. Consequently, $\boldsymbol{\Sigma}(t,t)$ is also non-singular.

Then, the proof idea of Proposition~2 in \textcite{baumeister_quantile-based_2024} carries over to the present setting.
Equation~(\ref{eq:xi_ats_consistent}) follows directly from the consistency of $\boldsymbol{\hat\Sigma}$ as $\Xi_{\textsc{ats}}(\mathbf{A},\boldsymbol{\hat\Sigma},t)$ $=\tr(\mathbf{A}\boldsymbol{\Sigma}(t,t)\mathbf{A}^\top)^{-1}$ is a continuous function of $\boldsymbol{\hat\Sigma}$.
Consequently, the term $\mathbf{A}\boldsymbol{\hat\Sigma}(t,t)\mathbf{A}^\top$ is also consistent for $\mathbf{A}\boldsymbol{\Sigma}(t,t)\mathbf{A}^\top$ and $(\mathbf{A}\boldsymbol{\hat\Sigma}(t,t)\mathbf{A}^\top)_0$ is consistent for $(\mathbf{A}\boldsymbol{\Sigma}(t,t)\mathbf{A}^\top)_0$.
The continuity of the Moore-Penrose Inverse follows from Theorem~4.2 of \textcite{rakocevic_continuity_1997}. 
The condition of Theorem~4.2, no rank jumps in $\mathbf{A}\boldsymbol{\hat\Sigma}(t,t)\mathbf{A}^\top$ respective $(\mathbf{A}\boldsymbol{\hat\Sigma}(t,t)\mathbf{A}^\top)_0$, is fulfilled by Assumption~\ref{ass:A3}.
From this, the Equations~(\ref{eq:xi_wts_consistent}) and (\ref{eq:xi_mats_consistent}) follow immediately.

\end{proof}

To investigate both, the asymptotic behaviour under the null $\mathcal{H}_0$ and under the alternative $\mathcal{H}_1$, recall that $\mathcal{H}_0$ and $\mathcal{H}_1$ can be written as 
\begin{align}
\mathcal{H}_0:\,\forall\, t\in\mathcal{T}:\,\eta_1(t)=\eta_2(t) &\Leftrightarrow \forall\, t\in\mathcal{T}:\,\mathbf{A}\boldsymbol{\mu}=\mathbf{0},\label{eq:hypothesis_null_A}\\
\mathcal{H}_1:\,\exists\, t\in\mathcal{T}:\,\eta_1(t)\not=\eta_2(t) &\Leftrightarrow \exists\, t\in\mathcal{T}:\,\mathbf{A}\boldsymbol{\mu}\not=\mathbf{0}\label{eq:hypothesis_alt_A}.
\end{align}

\paragraph{Asymptotic behaviour under the null hypothesis.}  
Assume $\mathcal{H}_0$.
Combining Proposition~\ref{thm:Aclt} and Proposition~\ref{thm:xi_consistent} with Slutzky`s Theorem leads to
\begin{align*}
&\left(T_n(\mathbf{A},\mathbf{x},t)\right)_{t\in\mathcal{T}}\\ =&\left(n\left(\mathbf{A}\boldsymbol{\hat\mu}(t)\right)^\top\Xi\left(\mathbf{A},\boldsymbol{\hat\Sigma},t\right)\mathbf{A}\boldsymbol{\hat\mu}(t)\right)_{t\in\mathcal{T}}\\
=&\left(\sqrt{n}\left(\mathbf{A}\boldsymbol{\hat\mu}(t)\right)^\top\Xi\left(\mathbf{A},\boldsymbol{\hat\Sigma},t\right)\sqrt{n}\mathbf{A}\boldsymbol{\hat\mu}(t)\right)_{t\in\mathcal{T}} \stackrel{d}{\longrightarrow}\left(\mathbf{Z}(t)^\top\,\Xi\left(\mathbf{A},\boldsymbol{\Sigma},t\right)\mathbf{Z}(t)\right)_{t\in\mathcal{T}}
\end{align*}
in $C(\mathcal{T})$.
Combined with the continuity of the functions $t\mapsto\int_\mathcal{T}\,\cdot\, dt$ and $t\mapsto\sup_{t\in\mathcal{T}}(\,\cdot\,)$ on $C(\mathcal{T})$ and the Continuous Mapping Theorem, this leads to the first assertion of Theorem~\ref{thm:distribution_test_statistics}.

\paragraph{Asymptotic behaviour under the alternative.} 
Now assume $\mathcal{H}_1$.
With the uniform consistency of $\{\boldsymbol{\mu}(t):t\in\mathcal{T}\}$ and Proposition~\ref{thm:xi_consistent}, it can be argued that 
\begin{align}
\frac{1}{n}I_n(\mathbf{A},\mathbf{x})&\stackrel{p}{\longrightarrow}\int_{\mathcal{T}}\left(\mathbf{A}\boldsymbol{\mu}(t)\right)^\top\Xi\left(\mathbf{A},\boldsymbol{\Sigma}, t\right)\mathbf{A}\boldsymbol{\mu}(t)dt, \label{eq:alternative_consistent_int}\\
\frac{1}{n}S_n(\mathbf{A},\mathbf{x})&\stackrel{p}{\longrightarrow}\sup_{t \in \mathcal{T}}\left\{\left(\mathbf{A}\boldsymbol{\mu}(t)\right)^\top\Xi\left(\mathbf{A},\boldsymbol{\Sigma},t\right)\mathbf{A}\boldsymbol{\mu}(t)\right\}\label{eq:alternative_consistent_sup}.	
\end{align}
Therefore, it remains to prove that the limits in Equations~(\ref{eq:alternative_consistent_int}) and~(\ref{eq:alternative_consistent_sup}) are positive under $\mathcal{H}_1$.
As Equation~(\ref{eq:hypothesis_alt_A}) holds for one $t\in\mathcal{T}$, it also holds in a neighbourhood $\mathcal{S}$ of $t$ due to the continuity of $t\mapsto\boldsymbol{\eta}(t)$.
In the following, we consider the different versions of $\Xi$ separately and start with $\Xi_{\textsc{ats}}(\mathbf{A},\boldsymbol{\hat\Sigma},t)=\tr(\mathbf{A}\boldsymbol{\hat\Sigma}(t,t)\mathbf{A}^\top)^{-1}$ for the \textsc{ats}.
Then, with Equation~(\ref{eq:hypothesis_alt_A}), it follows that for all $t\in\mathcal{S}$:
\begin{align*}
 \mathbf{A}\boldsymbol{\mu}\not=\mathbf{0}\,\Rightarrow\,\left(\mathbf{A}\boldsymbol{\mu}(t)\right)^\top\mathbf{A}\boldsymbol{\mu}(t)>0 \,\Rightarrow\,\frac{\left(\mathbf{A}\boldsymbol{\mu}(t)\right)^\top\mathbf{A}\boldsymbol{\mu}(t)}{\tr\left(\mathbf{A}\boldsymbol{\Sigma}(t,t)\mathbf{A}^\top\right)}>0.
\end{align*}
The last deduction is valid because of an always positive trace, which follows from Assumption~\ref{ass:A3}.
All in all, we decude
\begin{align*}
 \int_{\mathcal{S}}\frac{\left(\mathbf{A}\boldsymbol{\mu}(t)\right)^\top\mathbf{A}\boldsymbol{\mu}(t)}{\tr\left(\mathbf{A}\boldsymbol{\Sigma}(t,t)\mathbf{A}^\top\right)}dt >0,\;\sup_{t \in \mathcal{S}}\left\{\frac{\left(\mathbf{A}\boldsymbol{\mu}(t)\right)^\top\mathbf{A}\boldsymbol{\mu}(t)}{\tr\left(\mathbf{A}\boldsymbol{\Sigma}(t,t)\mathbf{A}^\top\right)}\right\} >0,
\end{align*}
which means that the assertion follows for the \textsc{ats}.
Next, we consider the  \textsc{wts} version $\Xi_{\textsc{wts}}(\mathbf{A},\boldsymbol{\hat\Sigma},t) =(\mathbf{A}\boldsymbol{\hat\Sigma}(t,t)\mathbf{A}^\top)^+$.
For an arbitrary $t\in\mathcal{S}$, the covariance matrix $\boldsymbol{\Sigma}(t,t)$ is non-singular as $\boldsymbol{\Gamma}(t,t)$ is non-singular by Assumption~\ref{ass:A3}.
Therefore, the square-root $\boldsymbol{\Sigma}^{1/2}(t,t)$ is also non-singular.
Choose a $\boldsymbol{\tilde\mu}(t)$ such that $\boldsymbol{\mu}(t)=\boldsymbol{\Sigma}^{1/2}\boldsymbol{\tilde\mu}(t)$.
Similarly to the proof of Theorem~2 in \textcite{ditzhaus_qanova_2021}, we can argue for every $t\in\mathcal{S}$
\begin{align*}
	\mathbf{A}\boldsymbol{\Sigma}^{1/2}(t,t)\left[\left(\mathbf{A}\boldsymbol{\Sigma}^{1/2}(t,t)\right)^+\mathbf{A}\boldsymbol{\mu}(t)\right]
	&=\mathbf{A}\boldsymbol{\Sigma}^{1/2}(t,t)\boldsymbol{\tilde\mu}(t)\\
	& = \mathbf{A} \boldsymbol{\Sigma}^{1/2}(t,t) \left(\boldsymbol{\Sigma}^{1/2}(t,t)\right)^+\boldsymbol{\mu}(t) = \mathbf{A}\boldsymbol{\mu}(t)\not  = \mathbf{0}.
\end{align*}
With the well-known properties of Moore--Penrose inverses \parencite[cf.][]{rao_generalized_1971}, it follows
\begin{align*}
\left(\mathbf{A}\boldsymbol{\mu}(t)\right)^\top\left(\mathbf{A}\boldsymbol{\Sigma}(t,t)\mathbf{A}^\top\right)^+\mathbf{A}\boldsymbol{\mu}(t)&=\left(\mathbf{A}\boldsymbol{\mu}(t)\right)^\top\left[\mathbf{A}\boldsymbol{\Sigma}^{1/2}(t,t)\left(\mathbf{A}\boldsymbol{\Sigma}^{1/2}(t,t)\right)^\top \right]^+\mathbf{A}\boldsymbol{\mu}(t)\\
&=\left(\mathbf{A}\boldsymbol{\mu}(t)\right)^\top\left[\left(\boldsymbol{\mathbf{A}\Sigma}^{1/2}(t,t)\right)^\top\right]^+\left(\mathbf{A}\boldsymbol{\Sigma}^{1/2}(t,t)\right)^+\mathbf{A}\boldsymbol{\mu}(t)\\
&=\left(\mathbf{A}\boldsymbol{\mu}(t)\right)^\top\left[\left(\boldsymbol{\mathbf{A}\Sigma}^{1/2}(t,t)\right)^+\right]^\top\left(\mathbf{A}\boldsymbol{\Sigma}^{1/2}(t,t)\right)^+\mathbf{A}\boldsymbol{\mu}(t)\\
&=\left[\left(\boldsymbol{\mathbf{A}\Sigma}^{1/2}(t,t)\right)^+\left(\mathbf{A}\boldsymbol{\mu}(t)\right)^\top\right]^\top\left(\mathbf{A}\boldsymbol{\Sigma}^{1/2}(t,t)\right)^+\mathbf{A}\boldsymbol{\mu}(t)\\
&>0
\end{align*}
for every $t\in\mathcal{S}$.
Consequently, it holds
\begin{align*}
\int_{\mathcal{S}}\left(\mathbf{A}\boldsymbol{\mu}(t)\right)^\top\left(\mathbf{A}\boldsymbol{\Sigma}\mathbf{A}^\top\right)^+\mathbf{A}\boldsymbol{\mu}(t)dt>0, \quad \sup_{t\in\mathcal{S}}\left\{\left(\mathbf{A}\boldsymbol{\mu}(t)\right)^\top\left(\mathbf{A}\boldsymbol{\Sigma}\mathbf{A}^\top\right)^+\mathbf{A}\boldsymbol{\mu}(t)\right\}>0,
\end{align*}
which ensures the assertion for \textsc{wts}.
The arguments can be chosen equivalently for \textsc{mats} with
$\Xi_{\textsc{mats}}(\mathbf{A},\boldsymbol{\hat\Sigma},t) = (\mathbf{A}\boldsymbol{\hat\Sigma}(t,t)\mathbf{A}^\top)_0^+$  \parencite[cf.][]{baumeister_quantile-based_2024}.		

\subsection{Proof of Theorem~\ref{thm:distribution_boot_test_statistic}}\label{sec:proof_distribution_boot_test_statistic}
The same resampling procedure is used in \textcite{munko_multiple_2025} and \textcite{baumeister_resampling-based_2026} to infer other multivariate functional models and the proof follows almost directly from their results. 
In particular, from Lemma~7 in \textcite{munko_multiple_2025} and the Continuous Mapping Theorem, a resampling version of Proposition~\ref{thm:Aclt} follows:
\begin{align}\label{eq:boot_Aclt}
\left(\sqrt{n}\mathbf{A}\left(\boldsymbol{\hat\mu}^\mathcal{P}(t)\right)\right)_{t\in\mathcal{T}}\stackrel{d}{\longrightarrow}\mathbf{Z}
\end{align}
in $C^2(\mathcal{T})$ in probability given the data, where $\mathbf{Z}$ is the Gaussian process from Proposition~\ref{thm:Aclt}. 
Moreover, Lemma~8 in \textcite{munko_multiple_2025} and Assumption~\ref{ass:A4} ensures the uniform consistency of the parametric bootstrap covariance estimator $\boldsymbol{\hat{\Sigma}}_{\mathcal{P}}$. 
With the same arguments as in the proof of Theorem~\ref{thm:distribution_test_statistics}, it follows with Equation~(\ref{eq:boot_Aclt}) and Slutzky`s Theorem that
\begin{align*}
\left(T_{n}(\mathbf{A},\mathbf{x}_\mathcal{P},t)\right)_{t\in\mathcal{T}} \stackrel{d}{\longrightarrow}\left(\mathbf{Z}(t)^\top\,\Xi\left(\mathbf{A},\boldsymbol{\Sigma},t\right)\mathbf{Z}(t)\right)_{t\in\mathcal{T}}
\end{align*}
in $C(\mathcal{T})$ in probability given the data. 
Therefore, the assertion follows with the Continuous Mapping Theorem.

\printbibliography

\end{document}